\documentclass[]{aastex701}

\usepackage{amsmath}
\usepackage{subcaption}
\usepackage{array}
\usepackage{longtable}
\usepackage{lipsum}
\usepackage{lineno}

\begin{document}

\title{Single vs. Binary Origin: The Diversity of Stripped-Envelope Supernova Remnants}

\author[orcid=0009-0006-1396-8397,sname='Kawashima']{Gaku Kawashima}
\affiliation{Department of Astronomy, Kyoto University, Sakyo-ku, Kyoto 606-8502, Japan}
\email[show]{kawashima@kusastro.kyoto-u.ac.jp}  

\author[orcid=0000-0002-2899-4241,sname='Lee']{Shiu-Hang Lee}
\affiliation{Department of Astronomy, Kyoto University, Sakyo-ku, Kyoto 606-8502, Japan}
\affiliation{Kavli Institute for the Physics and Mathematics of the Universe (WPI), The University of Tokyo, Kashiwa 277-8583, Japan}
\email{herman@kusastro.kyoto-u.ac.jp}

\author[orcid=0000-0003-2611-7269,sname='Maeda']{Keiichi Maeda}
\affiliation{Department of Astronomy, Kyoto University, Sakyo-ku, Kyoto 606-8502, Japan}
\email{keiichi.maeda@kusastro.kyoto-u.ac.jp}

\author[orcid=0000-0002-7507-8115,sname='Patnaude']{Daniel Patnaude}
\affiliation{Smithsonian Astrophysical Observatory, 60 Garden Street, Cambridge, MA 02138, USA}
\email{dpatnaude@cfa.harvard.edu}

\begin{abstract}
Core-collapse supernova remnants (CCSNRs) are crucial for understanding the final stages of massive star evolution, as they reflect the imprints of their progenitors' pre-explosion activities.
However, the evolution of CCSNRs, particularly those originating from progenitors with high mass-loss rates---known as stripped-envelope SNRs (SESNRs)---remains poorly understood.
This is largely due to the lack of comprehensive numerical models connecting progenitor stars to their remnants, especially in the context of binarity.
In this study, we perform self-consistent simulations of CCSNRs from both single and binary progenitors, utilizing mass-loss histories and supernova ejecta profiles directly derived from stellar evolution and explosion calculations.
Our models reveal significant differences in the circumstellar medium (CSM) structures between single and binary progenitors, which drive distinct SNR dynamics and spectral characteristics.
We find that binary-stripped progenitors tend to produce SNRs with more monotonic CSM profiles, resulting in smoother shock dynamics and less pronounced X-ray luminosity peaks compared to their single-star counterparts.
Additionally, we introduce a new characteristic timescale, $t_{\rm CSM}$, defined by the total mass lost by the progenitor.
This timescale effectively scales the evolutionary phases of CCSNRs in complex CSM environments, thereby facilitating the comparison of SESNRs.
Given that observed elemental abundances in SNRs reflect the nucleosynthesis yields of the progenitor, our results highlight the importance of considering the dynamical state of SNRs when interpreting observed abundances.
This work provides a fiducial framework for future observational and theoretical studies of CCSNRs, particularly regarding the impact of binary evolution.
\end{abstract}


\section{INTRODUCTION}\label{introduction}
Supernova remnants (SNRs) play a pivotal role as they act as drivers of high-energy materials, such as nucleosynthesized heavy elements \citep{Woosley1995THENUCLEOSYNTHESIS,Tsujimoto1995RelativeSMC,Arnett1996SupernovaePresent,Thielemann1996Core-CollapseEjecta,Chieffi2004EXPLOSIVEZ,Maeda2022StellarNucleosynthesis}, and as sites of cosmic ray acceleration \citep[e.g., diffusive shock acceleration:][]{Fermi1949OnRadiation, Drury1983AnPlasmas, Caprioli2010Non-linearBoundary, Caprioli2010ComparisonAcceleration}.
While supernovae (SNe) are broadly classified by their explosion mechanisms into Type Ia (thermonuclear) and core-collapse (CC) types, core-collapse supernova remnants (CCSNRs), which arise from the explosions of massive stars ($\gtrsim8M_{\odot}$), are especially important in the context of the lifecycle of the universe, as they connect stellar death to the birth of new stars via their high masses.

Since massive stars eject a considerable amount of their mass during their lifetimes, the evolution of CCSNRs is governed by their interaction with the surrounding circumstellar medium (CSM), which is shaped by the complex mass-loss history of the progenitor.
Some SNe originate from progenitors that have experienced extensive mass loss, stripping away their outer envelopes; these are referred to as stripped-envelope SNe (SESNe), observationally classified as Type IIb, Ib, and Ic supernovae based on their early-time spectra \citep{Filippenko1997OPTICALSUPERNOVAE,Yoon2015EvolutionaryProgenitors}.
Among the SNRs suggested to have such origins—sometimes called stripped-envelope SNRs (SESNRs)—Cassiopeia A \citep[e.g.,][]{Krause2008TheIIbb,Bamba2025MeasuringXRISM/Resolveb,Agarwal2025ChlorineRemnant} and RX J1713.7-3946 \citep[e.g.,][]{Katsuda2015EvidenceJ1713.7-3946,Tateishi2021PossibleJ1713.73946} are notable examples where thermal X-ray emission has been detected, providing clues to their progenitor properties.

However, a clear mapping between the observed diversity of SESNRs, explosion types, and their specific pre-explosion progenitor scenarios remains elusive, largely due to the lack of comprehensive numerical models that account for the effects of binarity.
Although a significant fraction of massive stars exist in binary systems and interact with companions before exploding as CCSNe \citep{Sana2012BinaryStars}, the systematic differences between SNRs from single and binary progenitors remain poorly understood.
Currently, identifying an SNR as originating from a binary system relies on a patchwork of indirect evidence, such as a low ejecta mass coupled with a high shock velocity, a shell-dominated morphology, or oxygen-rich abundances, which often lacks coherent theoretical support.
Observational constraints alone are insufficient because SNRs are primarily detected via shock-heated material, making it difficult to directly infer explosion details or progenitor properties without robust theoretical models.

Therefore, numerical simulations are crucial for interpreting observations and constructing a coherent theoretical framework.
While foundational numerical models \citep[][]{Chevalier1982SELF-SIMILARMEDIUM,Truelove1999EvolutionRemnantsb}
have successfully described the hydrodynamical evolution of SNRs in simplified environments guided by self-similar analytical solutions, they often do not fully capture the diversity arising from realistic, structured CSM environments.
Furthermore, not only the CSM but also the detailed SN ejecta profile is important for modeling SNRs, since the observation of thermal emission from ejecta provides elemental records of pre-explosion composition, which is a key for back-engineering the remnants.
Recent hydrodynamical simulations have begun to address these complexities.
While systematic surveys have explored the broad impact of steady wind mass loss on SNR evolution \citep{Patnaude2015AreRemnants,Patnaude2017TheRemnants, Jacovich2021ALoss}, other studies have focused on specific scenarios; for example, \citet{Yasuda2019TimeEnvironments,Yasuda2021ResurrectionRemnants,Yasuda2021DarkRemnants} examined SNRs evolving through CSM shaped by winds from multiple evolutionary stages.
However, these models rely on canonical values inferred from observations, lacking the intrinsic link between a progenitor's time-dependent mass-loss history and its final internal structure.
In fact, modern stellar evolution studies emphasize the necessity of a comprehensive, self-consistent model, demonstrating that single and binary progenitors differ significantly in both mass-loss histories and pre-supernova core structures with time, which in turn affect SN yields \citep[][]{Laplace2021DifferentStars,Vartanyan2021Binary-strippedProgenitorsc,Farmer2023NucleosynthesisStars}.

In this work, we perform a systematic study of CCSNRs, specifically focusing on the distinct evolutionary paths of single and binary stripped-envelope progenitors.
We construct consistent, end-to-end numerical simulation pipelines that directly link detailed mass-loss histories and SN ejecta derived from the stellar evolution calculations of \citet{Farmer2023NucleosynthesisStars} to the formation of wind-blown CSM and the subsequent SNR hydrodynamics.
By creating a comprehensive set of fiducial models and synthesizing their X-ray emission, we aim to establish a unified picture of CCSNR evolution and provide a theoretical basis for understanding their observed diversity.
While recent studies have explored end-to-end modeling connecting stellar evolution to macroscopic feedback from supernova remnants, including binary populations \citep[e.g.,][]{Fichtner2024ConnectingRemnants}, our study specifically focuses on the detailed internal hydrodynamics during the SNR evolution and the synthesis of their X-ray emission. This approach aims to facilitate direct comparisons with observable SNRs, filling the gap between theoretical models of large-scale environment interactions and the properties of actual observed objects.

This paper is organized as follows.
Section~\ref{methods} describes our numerical methods, including the stellar evolution input models and hydrodynamical simulation setup.
Section~\ref{results} presents our results, focusing on the dynamical and spectral differences between single and binary progenitor scenarios.
In Section~\ref{discussions}, we discuss the implications of our findings and compare them with observations.
Section~\ref{conclusions} summarizes our conclusions.

\section{METHODS}\label{methods}
\subsection{Progenitors and SNe}\label{progenitors}
This study builds on the stellar evolution and explosion models of \citet{Farmer2023NucleosynthesisStars}.
We utilize their single and binary star models which were evolved from the zero-age main sequence (ZAMS) to the SN explosion following core collapse using the MESA stellar evolution code \citep[version 12115;][]{Paxton2011ModulesMESA,Paxton2013ModulesStars,Paxton2015ModulesExplosions,Paxton2018ModulesExplosions,Paxton2019ModulesConservation,Jermyn2023ModulesInfrastructure}.
We adopt the mass-loss histories and stellar properties at the onset of core collapse, as well as the chemical yields immediately prior to shock breakout.

Our SNR code computes non-equilibrium ionization states for 30 elements, requiring distribution of stable isotopes as input.
We converted the 162 unstable isotopes from the MESA output into their stable counterparts using the Python library \texttt{radioactivedecay} \citep{Amaku2010DecayAlgebra}.
This library calculates the resulting abundances from radioactive isotopes at a specified time, based on the ICRP decay table \citep[][]{ICRP2008NuclearCalculations}.
For the SN yields, we set a decay time of 100 years, since our simulation computes SNR evolution ranging from a few years to thousands of years post-explosion, and most short-lived isotopes decay within this period.
The decay calculation is particularly important for elements like $^{56}$Ni, which decays to $^{56}$Fe via $^{56}$Co, significantly affecting the iron abundance in the ejecta.

Detailed modeling of the SN-to-SNR transition phase (seconds to years post-explosion) requires computationally expensive, high-resolution, three-dimensional magneto-hydrodynamic simulations \citep[e.g.,][]{Orlando2025TracingSimulationsc}.
However, for the purpose of studying the long-term dynamical evolution of the remnant, the power-law ejecta profile described by \citet{Truelove1999EvolutionRemnantsb} serves as a robust approximation consistent with observations.
The density profile is given by:
\begin{equation}
  \rho(r, t_{\rm init}) = \begin{cases} 
    \rho_c & (r \le r_c) \\
    \rho_c (r/r_c)^{-n_{\rm SN}} & (r_c < r \le r_{\rm ej}) 
  \end{cases}
\end{equation}
where $\rho_{\rm c}$, $r_{\rm c}$, and $r_{\rm ej}$ are the core density, core radius, and ejecta radius, respectively.
These values are determined to match the innermost CSM density derived in the following section, using the given $M_{\rm ej}$, canonical kinetic energy of $K_{\rm SN} = 10^{51}$ erg, and initial simulation time $t_{\rm init} = 3$ years post-explosion, assuming the ejecta is in a homologous expansion phase.
The explosion energy $10^{51}$ is consistent with the energy injected in the models of \citet{Farmer2023NucleosynthesisStars}, which implies that all explosion energy is already converted into kinetic energy $K_{\rm SN} \sim E_{\rm SN}$.
To minimize the number of free parameters, we set the power-law index to $n_{\rm SN}=11$ for all models, which is within the range expected for CCSNRs \citep{Matzner1999TheSupernovaeb}. We note that the RSG/stripped-envelope progenitors will lead to slightly steeper/shallower density gradients, but we expect the difference is negligible for the purpose of the present investigation.

Table~\ref{tab:models} lists the models analyzed in this study.
For each channel (single and binary), we select three $M_{\rm ZAMS}$ values to compare low, medium, and high $M_{\rm ZAMS}$ stars.
Following the context of \citet{Farmer2023NucleosynthesisStars}, note that only models that successfully exploded using criteria established in \citet{Ertl2020TheLoss} are included, as some models that have similar $M_{\rm ZAMS}$ failed to explode due to non-monotonic compactness behavior \citep[e.g.,][]{Sukhbold2014THECORES}.
The ejecta mass is calculated as
\begin{equation}
    M_{\rm ej} = M_{\rm ZAMS} - \Delta M_{\rm wind} - \Delta M_{\rm RLOF} - M_{\rm CCO},
\end{equation}
where $M_{\rm CCO}$ is the central compact object (CCO) mass.
This value does not account for possible fallback accretion.

As shown in Table~\ref{tab:models}, binary models generally possess lower ejecta masses than single-star models of the same $M_{\rm ZAMS}$ due to mass loss via Roche-Lobe Overflow (RLOF).
Since we assume a fixed explosion energy, this difference in ejecta mass significantly impacts the initial shock velocity and subsequent SNR dynamics (see below for details of the binary model setup).
Therefore, in Section~\ref{discussions}, we also compare models with similar ejecta masses to isolate the effects of binarity from simple mass scaling.

As discussed in the following sections, the progenitor models and their SN ejecta can be roughly categorized into three groups according to their initial masses and single/binary nature:
\begin{enumerate}
    \item With H envelope (Single 11 $M_{\odot}$)
    \item Single stripped stars (Single 26, 33 $M_{\odot}$)
    \item Binary stripped stars (Binary 11, 15, 20, 26, 33 $M_{\odot}$)
\end{enumerate}
Henceforth, we mainly focus on the single and binary 11 and 26 $M_{\odot}$ star models to clarify the qualitative differences arising from the distinct types of progenitors.

\begin{deluxetable*}{cccccccc}
\setlength{\tabcolsep}{4pt}
\renewcommand{\arraystretch}{1.2}
\tablecaption{Model Parameters \label{tab:models}}
\tablewidth{\textwidth}
\tablehead{
\colhead{Channel} & \colhead{$M_{\rm ZAMS}$} & \colhead{$\rm H_{\rm fin}$} & \colhead{$\rm He_{\rm fin}$} & \colhead{$\Delta M_{\rm wind}$} & \colhead{$\Delta M_{\rm RLOF}$} & \colhead{$M_{\rm CCO}$} & \colhead{$M_{\rm ej}$}
}
\startdata
Single & 11.00 & 3.82 & 2.98 & 1.64 & - & 1.53 & 7.83 \\
Single & 26.00 & 0.05 & 2.20 & 14.06 & - & 2.08 & 9.86 \\
Single & 33.00 & 0.00 & 1.56 & 18.43 & - & 2.22 & 12.35 \\
Binary & 11.00 & 0.00 & 1.16 & 0.70 & 7.12 & 1.36 & 1.82 \\
Binary & 15.00 & 0.00 & 1.33 & 1.54 & 8.67 & 1.64 & 3.15 \\
    Binary & 20.00 & 0.00 & 1.39 & 2.09 & 10.83 & 1.62 & 5.45 \\
Binary & 26.00 & 0.00 & 0.86 & 5.37 & 11.32 & 1.96 & 7.35 \\
Binary & 33.00 & 0.00 & 0.29 & 10.08 & 10.94 & 2.15 & 9.83 \\
\enddata
\tablecomments{All values are in units of $M_{\odot}$. $\rm H_{\rm fin}$, $\rm He_{\rm fin}$ are the mass of $\rm ^1H$, $\rm ^4He$ at core collapse.}
\end{deluxetable*}

\subsection{CSM Simulation Setup}\label{csm}
Most previous numerical studies of SNRs have relied on simplified CSM structures, such as steady wind profiles ($\rho \propto r^{-2}$), to approximate the environment \citep[e.g.,][]{Truelove1999EvolutionRemnantsb}.
However, time-varying mass-loss rates during different evolutionary phases create complex CSM structures that crucially affect the long-term evolution of the remnant \citep{Yasuda2021ResurrectionRemnants}.
In this work, we construct realistic CSM models by performing one-dimensional hydrodynamic simulations of stellar winds interacting with the interstellar medium (ISM), using time-dependent mass-loss rates and wind velocities obtained from MESA \citep["Dutch" scheme;][]{deJager1988MassDiagram.}. 
Note that the wind rate prescription carries some uncertainty; recent studies indicate that RSG mass-loss rates may be significantly lower than those given by the "Dutch" prescription \citep{Mauron2011ThePrescription, Renzo2017SystematicStars, Beasor2020ASupergiants}.

We employ the 1D hydrodynamics code VH-1 \citep{Blondin2001PULSARREMNANTSb,Blondin2001RAYLEIGH-TAYLORACCELERATIONb} to calculate the CSM structure.
The simulation grid consists of 2000 zones with a linear/logarithmic hybrid spacing, extending from $10^{14}$ cm to 1000 pc, ensuring numerical stability and minimizing boundary effects.
For this wind simulation, unlike the SNR calculations described below, we calculate only the bulk hydrodynamical evolution and omit detailed abundance tracking for computational efficiency.
Consistent with this approximation, radiative cooling is included using the scheme of \citet{Townsend2009AnSimulations} assuming Collisional Ionization Equilibrium (CIE) with solar metallicity, since the bulk of the pre-SN mass loss is represented by the H-rich envelope.
This cooling effect is crucial as the piled-up wind material forms a dense, geometrically thin shell.

Throughout stellar evolution, we adopt the escape velocity $V_{\rm escape}(t) = \sqrt{2GM_{1}(t)/R_{1}(t)}$ as the wind velocity $V_{\rm w}(t)$, where $M_{1}(t)$ and $R_{1}(t)$ are the stellar mass and radius at time $t$.
This simplification is used because detailed wind acceleration mechanisms for all evolutionary phases are not fully understood; other studies have proposed weaker wind velocity prescriptions \citep[e.g.,][]{Eldridge2006TheAfterglows}.

Stars in our binary models experience binary interaction during their lifetimes.
In the models of \citet{Farmer2023NucleosynthesisStars}, the initial binary mass ratio is set to $M_2/M_1=0.8$, and orbital periods are selected to satisfy the case B mass transfer condition without entering into a common-envelope phase \citep{Paczynski1967GravitationalBinaries}.
They simplified the evolution by treating the companion as a point mass until the end of core helium burning, after which the primary star's evolution continues without the companion, assuming no further interaction \citep[see also][]{Laplace2020TheProgenitors}.
They also assumed that any mass lost via RLOF is not returned to the primary star.
Given that the exact fraction of mass expelled from the system during RLOF is uncertain, we assume half is accreted by the companion and half is ejected, contributing to the CSM.
We note that our CSM calculation assumes one-dimensional spherical symmetry, whereas mass ejection during RLOF is likely a non-spherically symmetric process, possibly concentrated in the equatorial plane.
Our one-dimensional model thus represents a spherically-averaged scenario for this ejected material.
The mass-loss history via RLOF depends on the specific binary parameters selected by \citet{Farmer2023NucleosynthesisStars}.
Consequently, variations in the binary setup could produce a greater variety of CSM structures.
A detailed investigation of this binary parameter dependence is left for future work.

\begin{figure*}[htbp]
\begin{center}
\includegraphics[width=180mm]{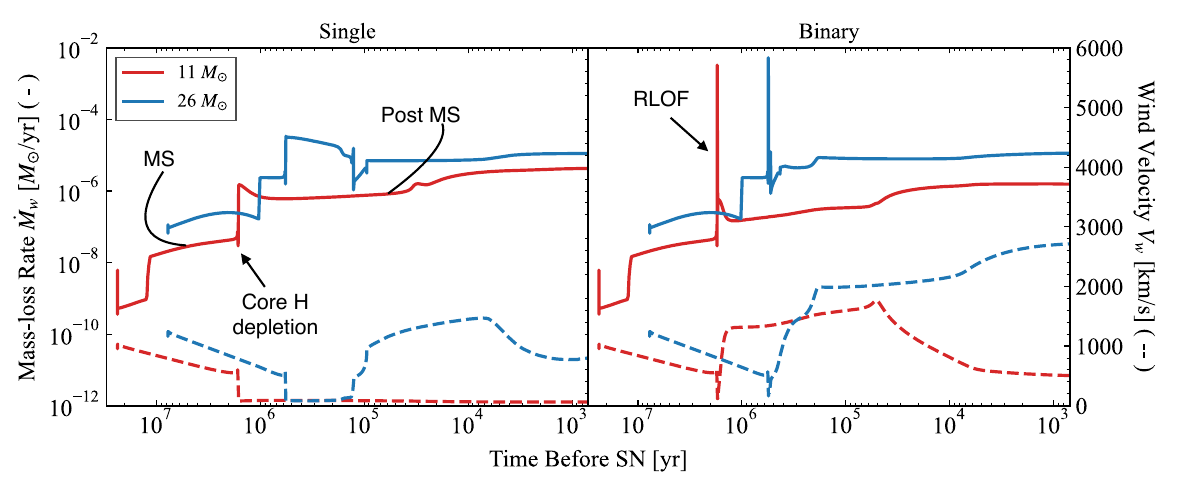}
\caption{Evolution of the wind parameters for the 11 and 26 $M_{\odot}$ models.
The solid lines show the time evolution of mass loss rates $\dot{M}(t)$ ($M_{\odot} \, \mathrm{yr}^{-1}$), and the dashed lines show the time evolution of wind velocity $V_{\rm w}(t)$ ($\mathrm{km\ s}^{-1}$).
The left and right panels represent the single star and binary star models, respectively.
The figure shows that the binary models have relatively stronger winds after RLOF compared to their single-star counterparts with the same $M_{\rm ZAMS}$.}
\label{fig:Mdot_Vw_prof}
\end{center}
\end{figure*}

Figure~\ref{fig:Mdot_Vw_prof} shows the time evolution of wind parameters for our models.
Initially, all models undergo a main sequence stage with low $\dot{M}$ and high $V_{\rm w}$.
After core hydrogen and helium exhaustion, their structures change significantly.
The balance of mass and radius leads to divergent wind velocities in single star models: heavier stars have faster WR winds, while lighter stars have slower RSG winds.
In binary cases, intense mass-loss via RLOF produces smaller progenitor masses at collapse, resulting in faster winds comparable to WR winds, even for lower-mass progenitors.
We assume the CSM metallicity matches the surface abundance at the time the wind was ejected.
We adopt the same decay procedure as for the ejecta, setting the decay time to the age of the wind material at the time of core collapse.
However, this detailed decay calculation is not significant for the CSM, as the ejected material consists mostly of stable elements.
Both wind and ISM temperatures are set to $T=10^4$ K, and the ISM density is $n_{\rm ISM}=1.0$ cm$^{-3}$.
We adopt a relatively high ISM density of $n_{\rm ISM}=1.0$ cm$^{-3}$ to represent a typical dense environment surrounding massive star-forming regions. While we acknowledge this is at the high end of typical ISM densities, this work focuses on establishing a fiducial model rather than performing a parameter study of the ambient medium.

Since we aim to track the overall evolution of the SNRs, we do not focus on resolving the final few $10^3$ years until core-collapse.
The mass ejected during the final $10^3$ years pre-explosion is assumed to be $\sim 10^{-3} M_\odot$, which is negligible for our SNR simulations.
Regarding our models in this study, this simplification is reasonable since drastic mass-loss events such as eruptions are not included (except for RLOF), and the rapid rise in mass-loss rate at the final moment \citep{Quataert2012Wave-drivenSupernovae} does not eject a considerable amount of mass in our models.
Note that this simplification may carry the risk of overlooking features in SNRs at very early stage, thus our models may not be directly comparable to such youngest SNRs.

\subsection{SNR Simulation Setup}\label{snr}
We perform one-dimensional hydrodynamic simulations in spherical coordinates using the Eulerian version of the ChN code \citep[e.g.,][]{Court2024DoNebulae}.
This code couples hydrodynamics (VH-1), time-dependent non-equilibrium ionization (NEI), and radiative cooling \citep[see also][]{Patnaude2017TheRemnants,Lee2012ARemnant,Lee2013AJ0852.0-4622,Lee2014X-rayA,Lee2015MODELINGSHOCKS,Jacovich2021ALoss}.
Radiative cooling is calculated using the exact integration scheme described by \citet{Townsend2009AnSimulations}. We adopt a fixed cooling function $\Lambda(T)$ assuming Collisional Ionization Equilibrium (CIE) with solar abundances, based on the tabulated values from \citet{Sutherland1993COOLINGPLASMAS}. We do not include spatially resolved elemental abundances or time-dependent ionization states in the calculation of the cooling term.
Our code employs a high-accuracy shock-capturing method that scans the grid inward and outward from the boundaries to precisely determine the positions of the forward and reverse shocks, identified by steep gradients in pressure and velocity at every timestep.
The simulation is initialized at 3 years post-explosion.
The CSM profile constructed in Section \ref{csm} is mapped onto the grid first, followed by the placement of the ejecta structure described in Section \ref{progenitors} to match the CSM.
We evolve the system until the reverse shock reaches the innermost simulation grid.
To avoid complications from effects such as pulsar winds, we do not include central compact objects in our simulations.

For the simulation setup, we adjust the box size individually to cover the outermost CSM shell for each model.
We used a total of 12000 zones with a linear/logarithmic hybrid spacing to resolve both the inner structure up to the outer shell, allocating about 1000 zones specifically to cover the ejecta with linear grid spacing.
We also prepared 12000 Lagrangian tracer particles to track actual properties in the shocked material.
A critical parameter for X-ray emission modeling is the electron temperature ($T_{\rm e}$), which may differ from the ion temperature ($T_{\rm i}$) in collisionless shocks.
In this work, we assume no anomalous collisionless electron heating at the shock front.
Consequently, the post-shock electron temperature is initialized strictly by the mass ratio relation $T_{\rm e} = T_{\rm i} (m_{\rm e}/m_{\rm i})$ where $m_{\rm e}$ and $m_{\rm i}$ are the electron and ion masses, respectively.
The subsequent evolution of $T_{\rm e}$ is governed by Coulomb collisions with ions and adiabatic expansion/compression and the radiative cooling.
This assumption corresponds to the minimal heating scenario ($\beta \approx \beta_{\rm min}$) discussed in \citet{Ghavamian2007ASHOCKS}, which is consistent with high shock velocity ($\gtrsim 2000$ km s$^{-1}$) SNRs.
For the calculation of NEI and X-ray spectra, we track the ionization states of 30 elements.
Synthetic X-ray spectra are generated using the SOXS X-ray modeling software \citep{ZuHone2023SOXS:Sources}.

\section{RESULTS}\label{results}
\subsection{CSM}\label{csm_results}
\begin{figure*}[htbp]
\begin{center}
\includegraphics[width=180mm]{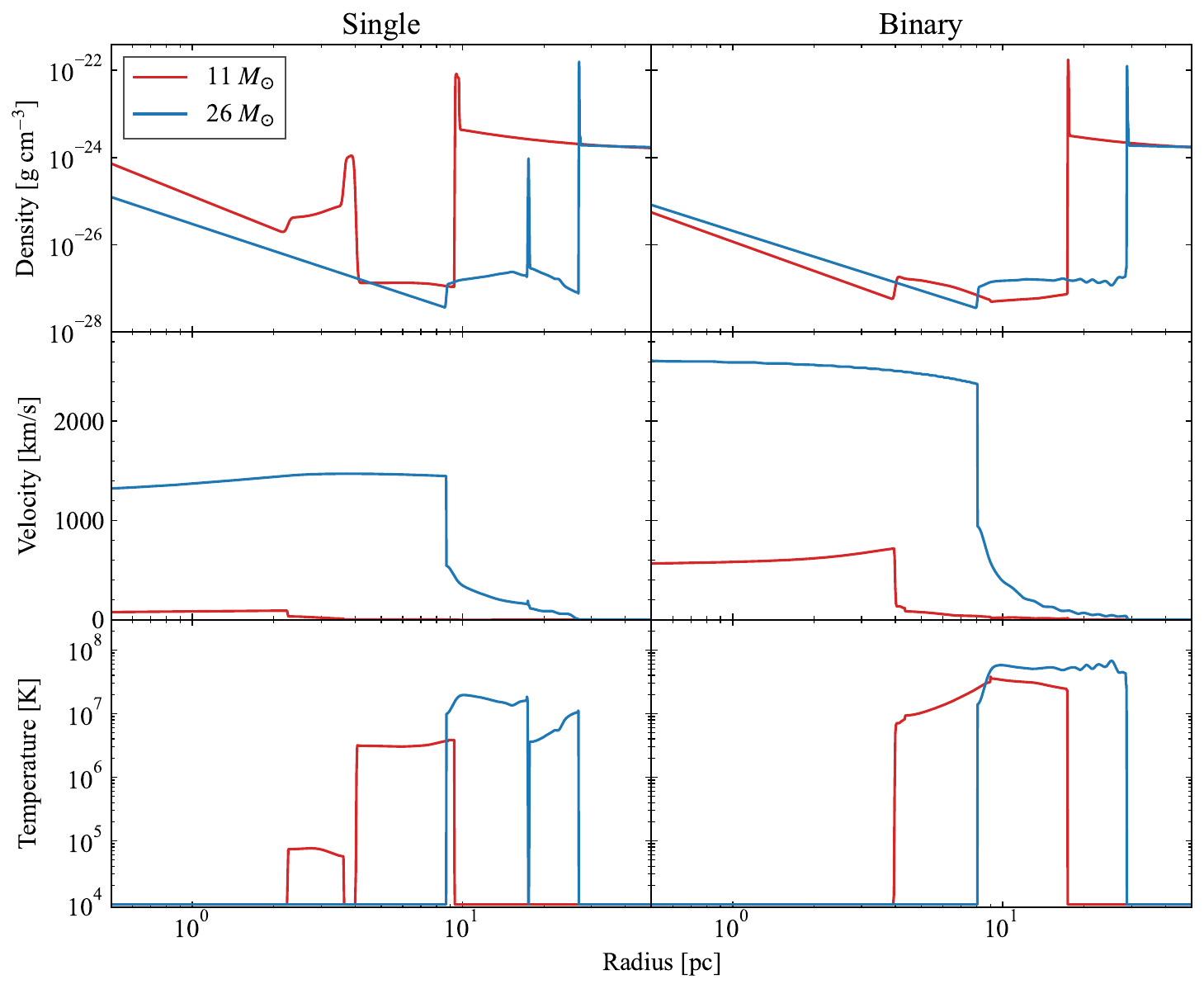}
\caption{CSM profiles of our models.
The top panel shows the gas density, the middle panel shows the gas velocity, and the bottom panel shows the gas temperature, as a function of radius.}
\label{fig:csm_prof}
\end{center}
\end{figure*}

Figure~\ref{fig:csm_prof} shows the resulting CSM density, velocity, and temperature profiles at the moment of core collapse.
Several general characteristic regions are visible: from outer to inner radius, there is a substantially uniform ISM region (which exhibits a smooth density gradient just outside the shell), a dense shell of swept-up ISM and CSM, a hot, low-density bubble, and an innermost region shaped by the most recent, freely expanding wind.
Notably, single star models exhibit internal shell structures within the bubble, while binary star models do not, resulting in more monotonic CSM profiles.

\begin{figure*}[htbp]
\begin{center}
\includegraphics[width=180mm]{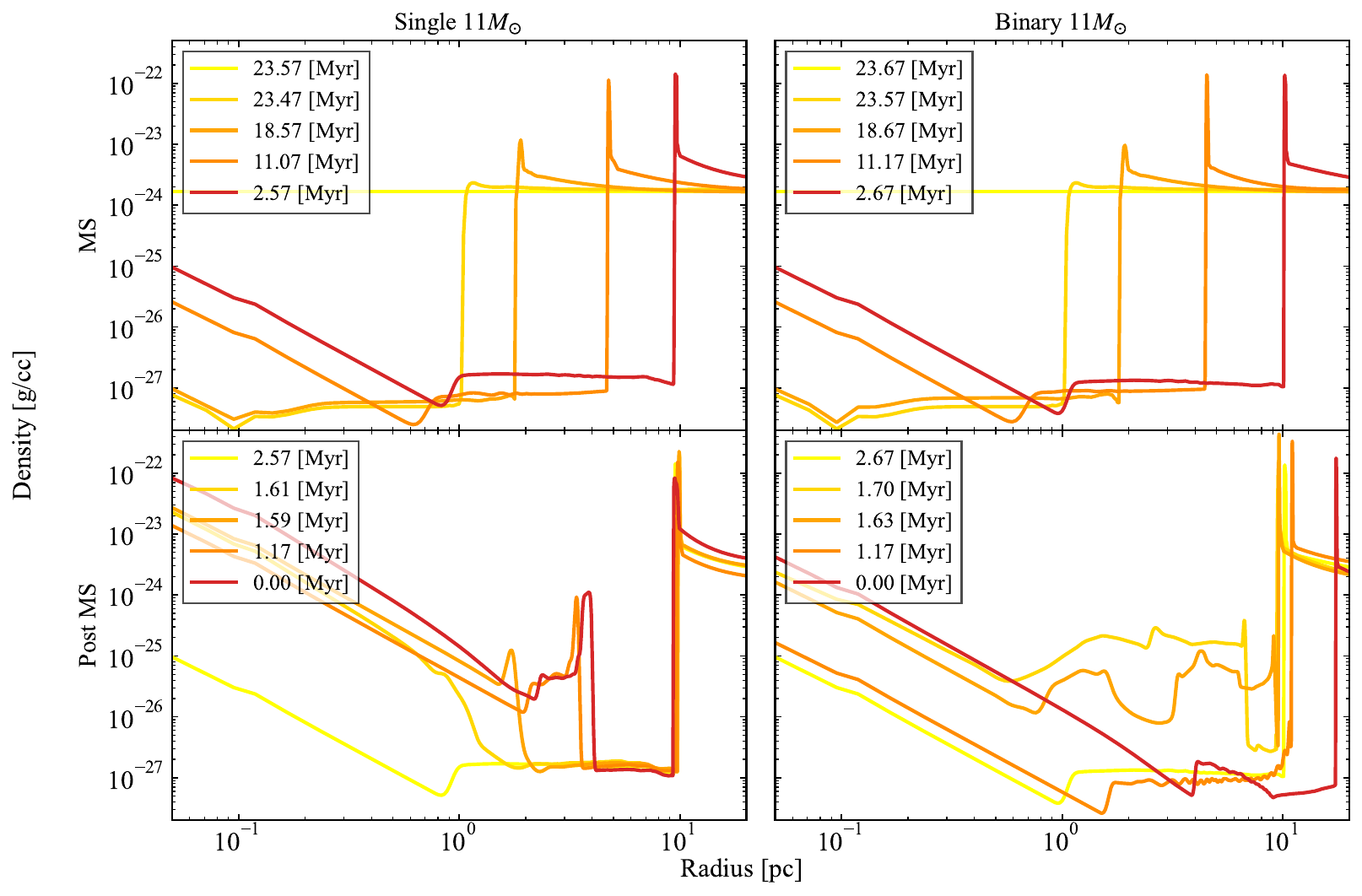}
    \caption{CSM density evolution of our 11 $M_{\odot}$ models. The left panel shows the single star model, and the right panel shows the binary star model. Note that the timestamps shown in the legend represent the "look-back time" before the SN, consistent with Figure~\ref{fig:Mdot_Vw_prof}. This figure shows how the stronger winds in the binary models blow away the mass ejected via RLOF.}
\label{fig:csm_animation}
\end{center}
\end{figure*}

Figure~\ref{fig:csm_animation} illustrates the dynamical formation of the CSM for the 11 $M_{\odot}$ models.
The size of the wind-blown bubble is largely determined by the integrated power of the stellar winds \citep{Weaver1977InterstellarEvolution.}.
The strength of the wind—defined by the wind velocity and mass-loss rate—plays a crucial role in shaping the inner structure.
Consequently, strong winds from binary-stripped stars (or highly stripped single stars like the 33 $M_{\odot}$ model) sweep previously formed inner structures outward.
This sweeping effect explains the tendency for binary star models to have more monotonic CSM profiles compared to their single-star counterparts.

Table~\ref{tab:csm_mass} represents the inner structure of the CSM. The outer region starting from the CSM+ISM shell, acts as a fixed boundary for the system due to the extremely high mass contained within the shell, as can be seen in the following sections.

\begin{deluxetable*}{ccccccccccc}
\setlength{\tabcolsep}{4pt}
\renewcommand{\arraystretch}{1.2}
\tablecaption{CSM Properties by Region and Mass Ratios \label{tab:csm_mass}}
\tablewidth{\textwidth}
\tablehead{
\colhead{Channel} & \colhead{$M_{\rm ZAMS}$} & \colhead{$R_{\rm bubble}$} & \colhead{$R_{\rm shell}$} & \colhead{$M_{\rm wind}$} & \colhead{$M_{\rm bubble}$} & \colhead{$M_{\rm outer}$} & \multicolumn{2}{c}{$M/M^{\rm loss}_{\rm CSM}$ (\%)} & \multicolumn{2}{c}{$M/M_{\rm ej}$ (\%)} \\
\colhead{} & \colhead{($M_{\odot}$)} & \colhead{(pc)} & \colhead{(pc)} & \colhead{($M_{\odot}$)} & \colhead{($M_{\odot}$)} & \colhead{($M_{\odot}$)} & \colhead{wind} & \colhead{bubble} & \colhead{wind} & \colhead{bubble}
}
\startdata
Single & 11.00 & 2.26 & 9.48 & 0.07 & 1.09 & 8.9e+07 & 4.1 & 66.8 & 0.9 & 14.0 \\
Single & 26.00 & 8.71 & 27.01 & 0.05 & 7.75 & 1.0e+08 & 0.3 & 55.2 & 0.5 & 78.6 \\
Single & 33.00 & 18.53 & 35.98 & 0.18 & 12.51 & 1.0e+08 & 1.0 & 67.9 & 1.4 & 101.3 \\
Binary & 11.00 & 3.98 & 17.45 & 0.01 & 0.62 & 9.3e+07 & 0.2 & 14.6 & 0.4 & 34.2 \\
Binary & 15.00 & 6.38 & 23.00 & 0.01 & 0.62 & 1.0e+08 & 0.2 & 10.6 & 0.4 & 19.8 \\
Binary & 20.00 & 6.64 & 21.43 & 0.02 & 2.65 & 9.9e+07 & 0.2 & 35.2 & 0.3 & 48.6 \\
Binary & 26.00 & 8.07 & 28.84 & 0.03 & 3.28 & 1.0e+08 & 0.3 & 29.7 & 0.4 & 44.6 \\
Binary & 33.00 & 10.95 & 36.91 & 0.06 & 5.30 & 1.0e+08 & 0.4 & 34.1 & 0.6 & 53.9 \\
\enddata
\tablecomments{$R_{\rm bubble}$ and $R_{\rm shell}$ are the inner boundaries of the hot-bubble region and dense shell.
$M_{\rm wind}$, $M_{\rm bubble}$, and $M_{\rm outer}$ are the integrated gas masses.
$M_{\rm outer}$ represents the total integrated mass from $R_{\rm shell}$ to the outer boundary of the simulation domain.
Because this region includes the massive swept-up shell and the ambient ISM, its extremely large mass effectively acts as a fixed boundary in the hydrodynamical calculations.
Ratios are shown as percentages relative to the CSM threshold mass ($M^{\rm loss}_{\rm CSM}$) and the ejecta mass ($M_{\rm ej}$).}
\end{deluxetable*}

\subsection{SNR}\label{SNR_results}

\begin{figure*}[htbp]
    \begin{center}
    \includegraphics[width=180mm]{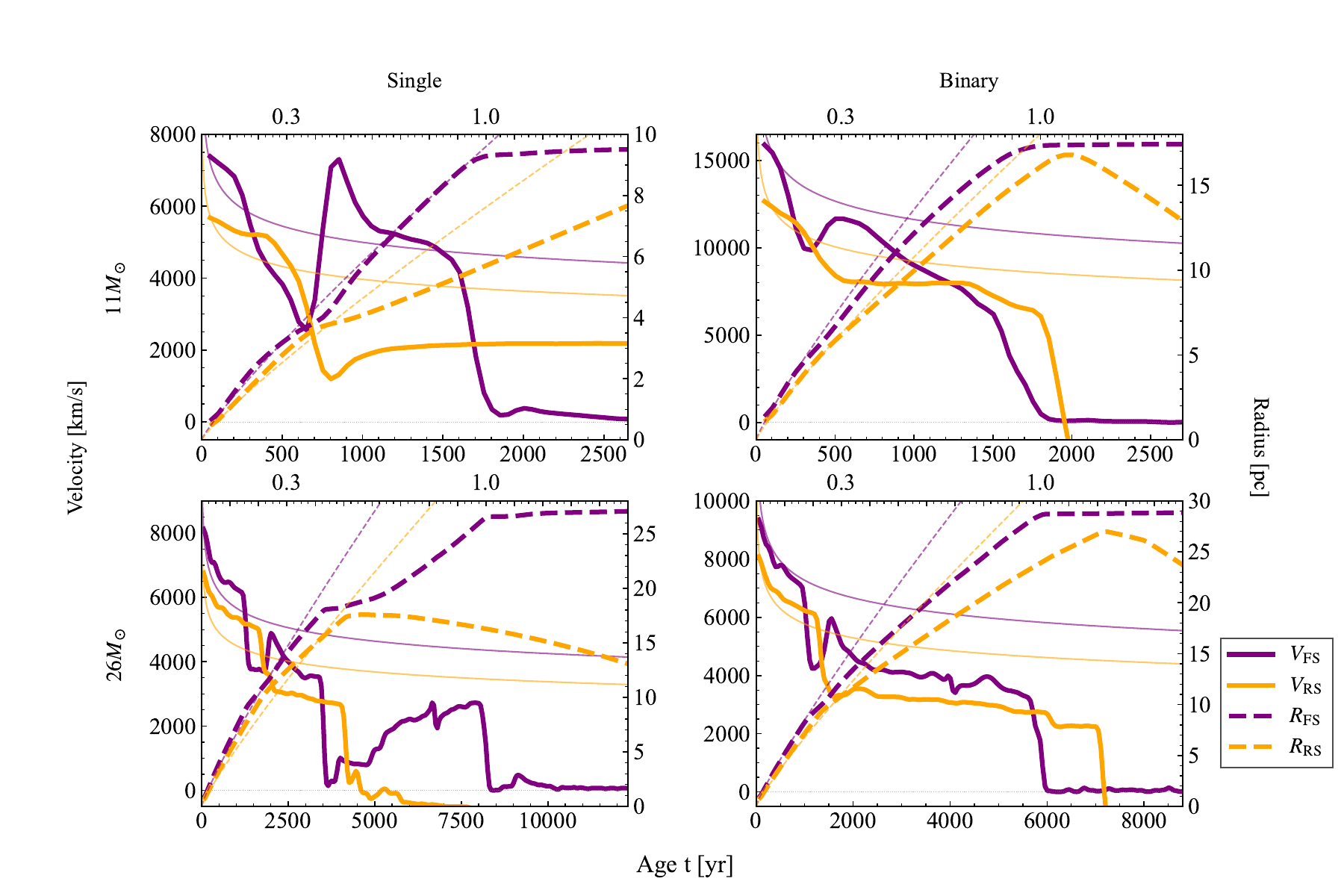}
    \caption{The shock radius and velocity evolution of our 11, 26 $M_{\odot}$ models.
    The purple line shows the radius of the forward shock, and the orange line shows the reverse shock, with solid lines representing the velocity and dashed lines representing the radius. Note that all velocities are in the observer's frame. Labels at the top of each panel represent the time scaled by $t_{\rm CSM}$ (defined below). Thin lines represent the analytical solutions from \citet{Truelove1999EvolutionRemnantsb}.
    The gradual deviation from the analytical solutions suggests that SNRs evolving within complex CSM cannot be accurately described by a simplified evolutionary model.}
    \label{fig:snr_dynamics}
    \end{center}
\end{figure*}

Figure~\ref{fig:snr_dynamics} displays the dynamical evolution of our SNR models.
In general, all models experience a significant deceleration upon reaching the outermost shell.
Due to the low-density interior of the wind-blown bubble, the reverse shock (RS) closely follows the forward shock (FS) for a significant portion of the early evolution.
A rapid decline in FS velocity marks the interaction with the dense outer shell.
After the FS collides with the dense outermost shell, a reflected shock is generated, which propagates inward and accelerates the RS toward the center.

The evolution of the FS and RS in our models follows classical SNR theory \citep{Truelove1999EvolutionRemnantsb} during the early stages ($\sim 10^3$ yr).
However, once the shocks break out of the initial power-law wind region, the evolution becomes complex and cannot be described analytically.
Traditionally, the "Sedov time" ($t_{\rm sedov}$), defined as the time when the swept-up CSM mass equals the ejecta mass ($M^{\rm sh}_{\rm CSM} = M^{\rm init}_{\rm ej}$), is adopted as a key timescale separating the SNR evolutionary phases, from the ejecta-dominated to the Sedov-Taylor phase \citep{Vink2020PhysicsRemnants}, which corresponds to the reversal of the RS.
However, our simulations reveal that for SNRs evolving in wind-blown CSM, $t_{\rm sedov}$ is not an adequate indicator for tracking all evolutionary phases originating from environmental complexities.

\begin{figure}[htbp]
\begin{center}
\includegraphics[width=180mm]{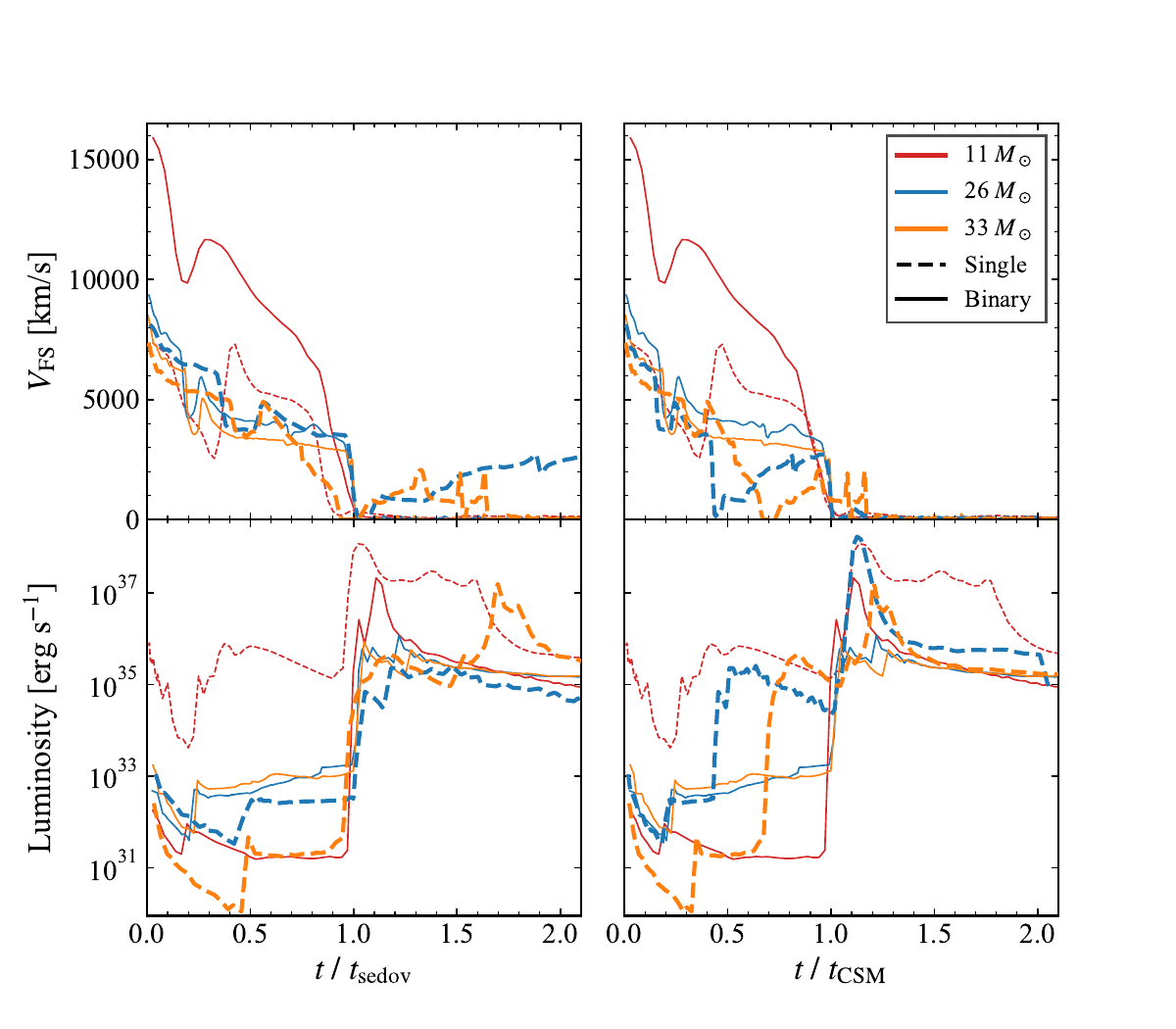}
\caption{Forward shock velocity (top) and total 0.3-12.0 keV luminosity (bottom) evolution of our models.
In the left panels, the SNR age (x-axis) is scaled by the Sedov time ($t_{\rm sedov}$), defined as when the swept-up CSM mass equals the ejecta mass ($M^{\rm sh}_{\rm CSM} = M^{\rm init}_{\rm ej}$).
In the right panels, the age is scaled by our new characteristic time, $t_{\rm CSM}$, defined as when the swept-up CSM mass equals the total mass lost by the progenitor ($M^{\rm sh}_{\rm CSM} = M^{\rm loss}_{\rm CSM}$).
The panels show that models with significant mass within the bubble (e.g., the single 26 and 33 $M_{\odot}$ models), are not aligned with the other models when scaled by $t_{\rm sedov}$, whereas scaling by $t_{\rm CSM}$ aligns them effectively. This suggests that our new scaling is useful for comparing SESNRs that have experienced considerable mass loss.}
\label{fig:fig_vs_sedov}
\end{center}
\end{figure}

Figure~\ref{fig:fig_vs_sedov} presents the time evolution of the forward shock velocity and total luminosity of thermal X-ray in our models, with time scaled by both the Sedov time ($t_{\rm sedov}$) and a new characteristic time, $t_{\rm CSM}$, defined as the age when the swept-up CSM mass equals the total mass lost by the progenitor via winds and RLOF ($M^{\rm sh}_{\rm CSM} = M^{\rm loss}_{\rm CSM}$).
This corresponds to the moment when the FS breaks into the dense shell.
The left panel shows the limitations of using $t_{\rm sedov}$ as a scaling parameter for our models; models with different ejecta masses reach this time at vastly different physical ages and evolutionary states.
This discrepancy is particularly evident in the X-ray light curves, where the peak luminosity occurs at widely varying fractions of $t_{\rm sedov}$.
In contrast, scaling the ages by our new reference time, $t_{\rm CSM}$ (right panel), aligns the evolutionary phases of the different models much more effectively.
This scaling reveals three distinct phases:
\begin{itemize}
    \item Phase 1: Expansion in the inner wind region ($0.0 \lesssim t/t_{\rm CSM} \lesssim 0.3$)
    \item Phase 2: Expansion through the hot, low-density bubble ($0.3 \lesssim t/t_{\rm CSM} \lesssim 1.0$)
    \item Phase 3: Interaction with the dense outer shell ($1.0 < t/t_{\rm CSM}$)
\end{itemize}
Our results suggest that most of the diversity among the models manifests during the second phase.
Interestingly, despite the differences in ejecta mass and wind properties, the duration of the first phase is remarkably similar across most models ($0.0 \lesssim t/t_{\rm CSM} \lesssim 0.3$), and the luminosity evolution follows a similar trend.
The main exception is the Single 11 $M_{\odot}$ model, whose slow RSG wind creates a denser inner CSM.
From the light curves, it is evident that the luminosity jump after the FS hits the outer shell is suppressed in binary models.
This suppression arises from the difference in previously shocked CSM mass; single models propagate through a denser bubble with an inner shell, whereas binary models expand with minimal interference.
Our introduction of $t_{\rm CSM}$ does not supersede the utility of $t_{\rm sedov}$; $t_{\rm sedov}$ remains an important indicator of RS dynamics in our models. However, for shell-like SESNRs where the thermal emission from the ejecta is dim—or where the emission is dominated by non-thermal radiation from cosmic rays accelerated at the FS, such as in Vela Jr.—$t_{\rm CSM}$ may provide a more consistent evolutionary baseline. The broader applicability of this new indicator will be explored in future studies.

\begin{figure*}[htbp]
\begin{center}
\includegraphics[width=180mm]{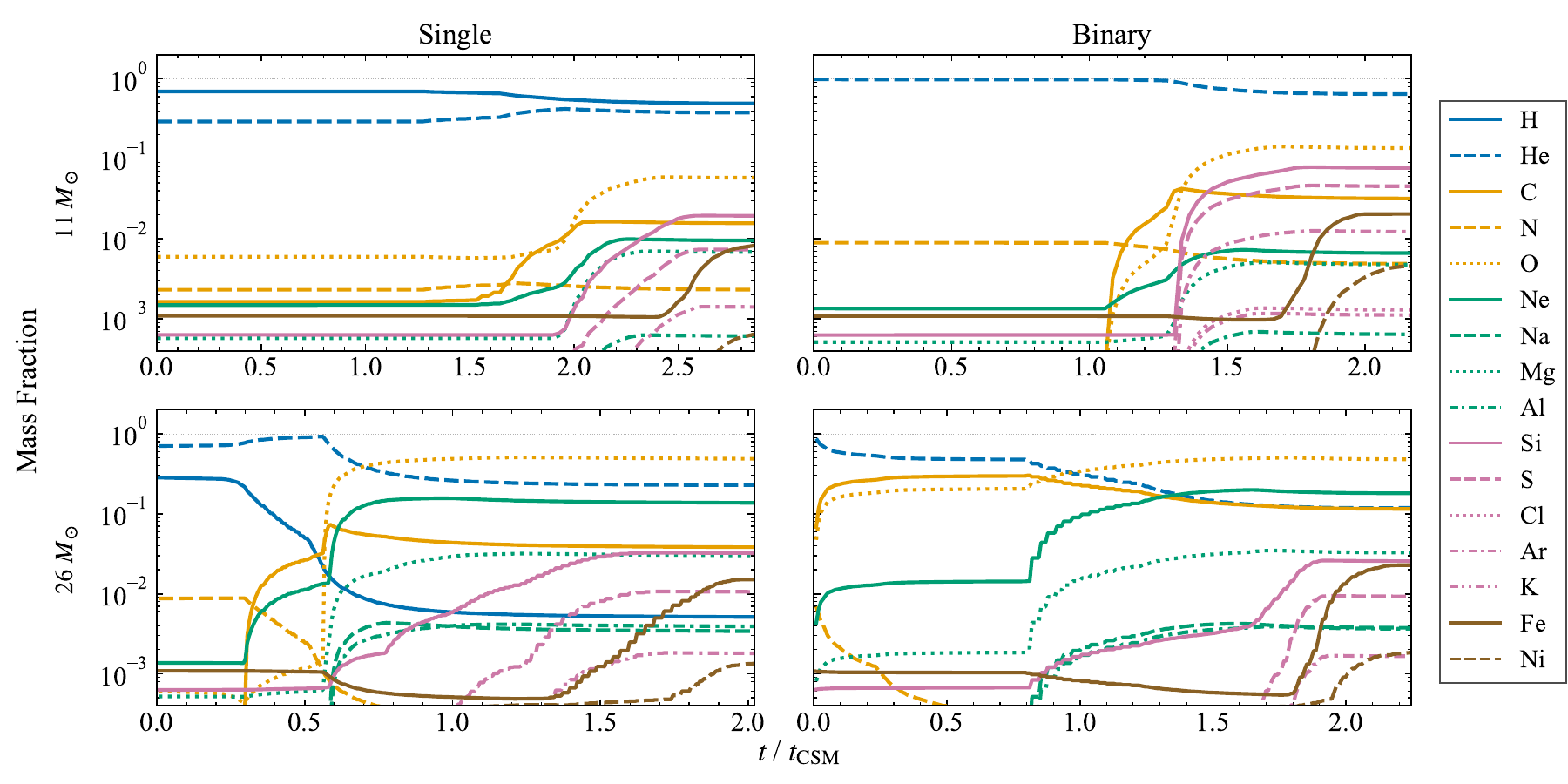}
\caption{Time evolution of the mass fractions of different elements of interest in the shocked ejecta in our 11 and 26 $M_{\odot}$ models. The exact values of shocked mass at each phases are listed in appendix.}
\label{fig:snr_shockedejecta}
\end{center}
\end{figure*}

\begin{figure*}[htbp]
\begin{center}
\includegraphics[width=180mm]{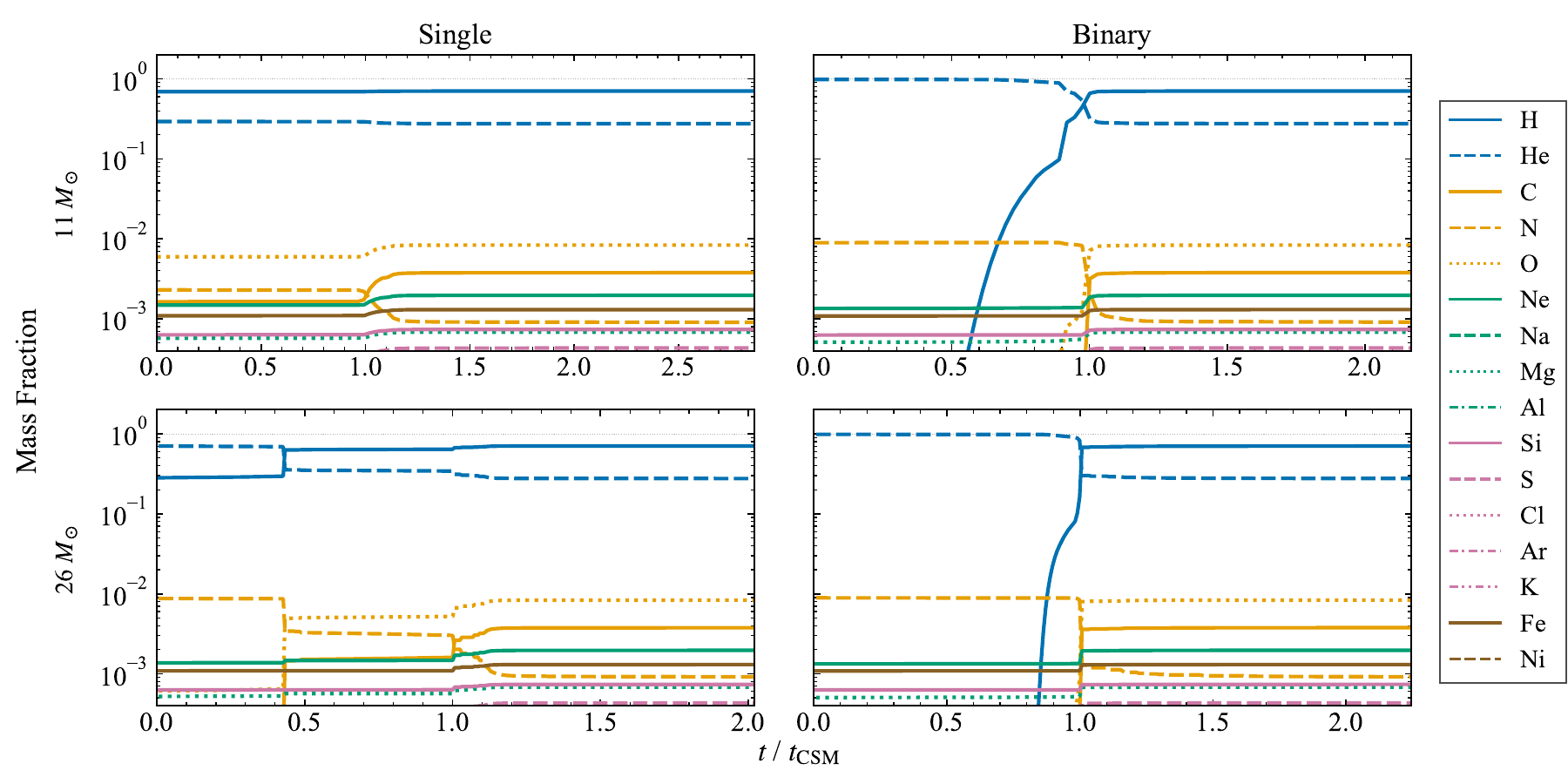}
\caption{Same as Figure~\ref{fig:snr_shockedejecta}, but for the shocked CSM.}
\label{fig:snr_shockedcsm}
\end{center}
\end{figure*}

Figures~\ref{fig:snr_shockedejecta} and \ref{fig:snr_shockedcsm} show the mass fractions of major elements in the RS-heated ejecta and FS-heated CSM, respectively.
The evolution of the shocked composition can be broadly categorized into five groups based on nucleosynthesis origins:
\begin{itemize}
    \item Envelope products: H and He.
    \item Hydrostatic burning products: C, N, O.
    \item Lighter IMEs (Intermediate Mass Elements): Ne, Na, Mg, and Al.
    \item Heavier IMEs: Si, S, Cl, Ar and K.
    \item IGEs (Iron-group elements): Fe and Ni.
\end{itemize}
As shown in the figures, the shocked composition of the SNR changes significantly over time as the reverse shock propagates through the stratified ejecta.
This implies that direct comparisons between observed SNR abundances and progenitor SN yields can be misleading without accounting for the remnant's dynamical state and which parts of the ejecta have been shocked.

\begin{figure*}[htbp]
\begin{center}
\includegraphics[width=180mm]{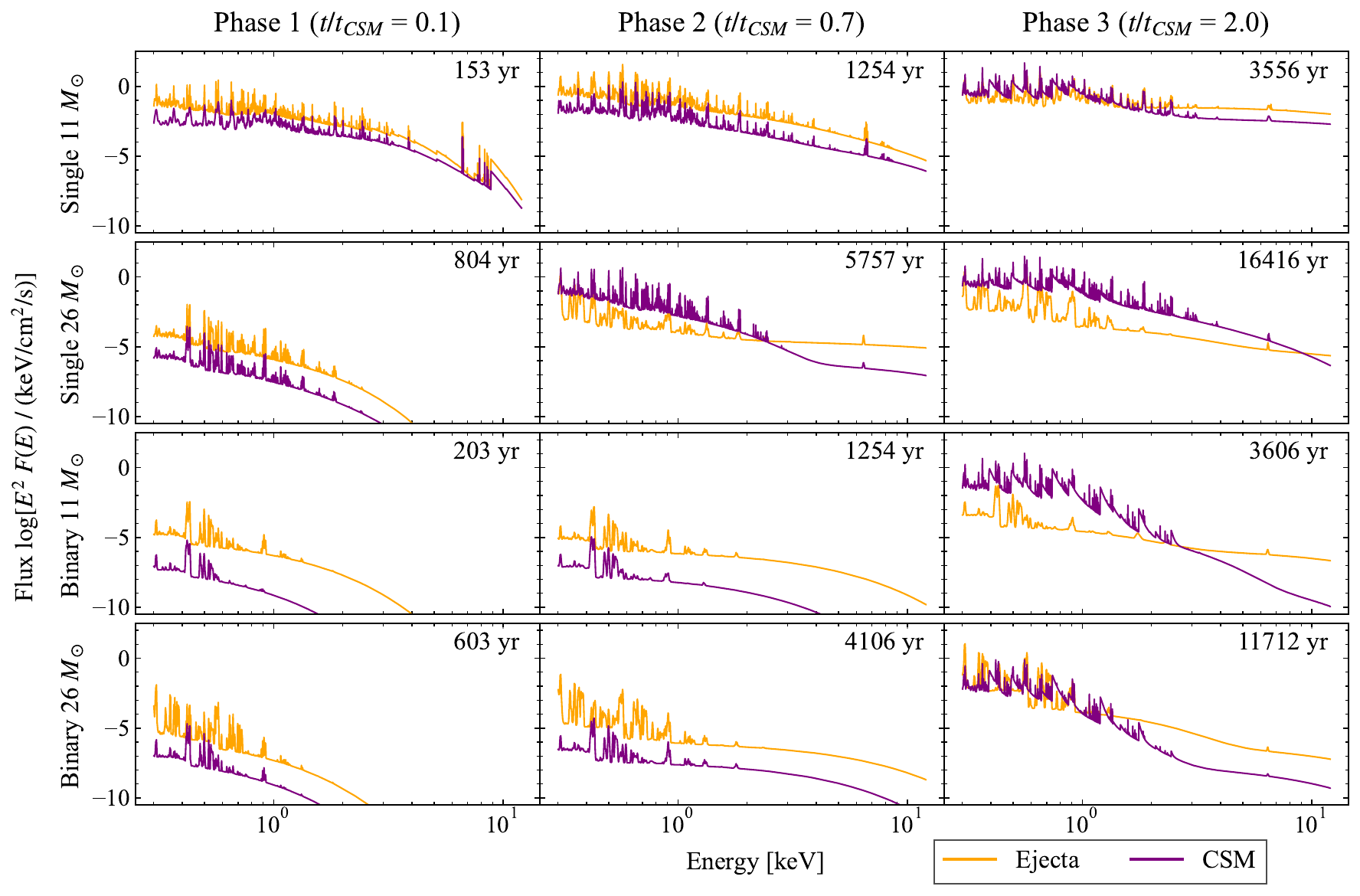}
\caption{Simulated X-ray spectra from our 11, 26 $M_{\odot}$ models at selected reference times based on our CSM mass-based scaling.
Orange and purple lines represent emissions from RS-heated ejecta, and FS-heated CSM, respectively.}
\label{fig:snr_emission}
\end{center}
\end{figure*}

Figure~\ref{fig:snr_emission} presents the time evolution of simulated X-ray spectra from our models at selected reference times scaled by $t_{\rm CSM}$.
A general trend observed in all stripped models (i.e., those that have lost a considerable amount of their H/He envelope mass, excluding the Single 11 $M_{\odot}$ model) is that the ejecta emission dominates the total X-ray emission at early phases, although it is relatively faint.
The faintness of SESNRs at early stages is also discussed in the previous study in the context of non-thermal emission \citep{Yasuda2021ResurrectionRemnants}.
This inherent faintness may explain the scarcity of detected binary SNRs.

Subsequently, CSM emission overwhelms the ejecta emission after the forward shock hits the dense outer shell ($t_{\rm CSM} \sim 1.0$).
We also observe a double component in the CSM continuum emission after the shock hits the shell.
This feature indicates the presence of two distinct temperature components in the shocked CSM: one from the material shocked before hitting the shell and another from the material currently being shocked at the dense shell.

Furthermore, most models in the late stage ($t_{\rm CSM} \gtrsim 1.0$) and the single 11 $M_{\odot}$ model in the early stage ($t_{\rm CSM} \sim 0.1$) show prominent cliff-like structures in the continuum, which are known as recombination edges \citep[][]{Yamaguchi2009DiscoverySuzaku,Ozawa2009SuzakuW49B}.
This is a signature of rapidly cooling, over-ionized plasma, which forms as the high-velocity shock propagates into the dense CSM \citep{Vink2011SupernovaPerspective,Moriya2012ProgenitorsRemnants}.

\begin{figure}[htbp]
\begin{center}
\includegraphics[width=180mm]{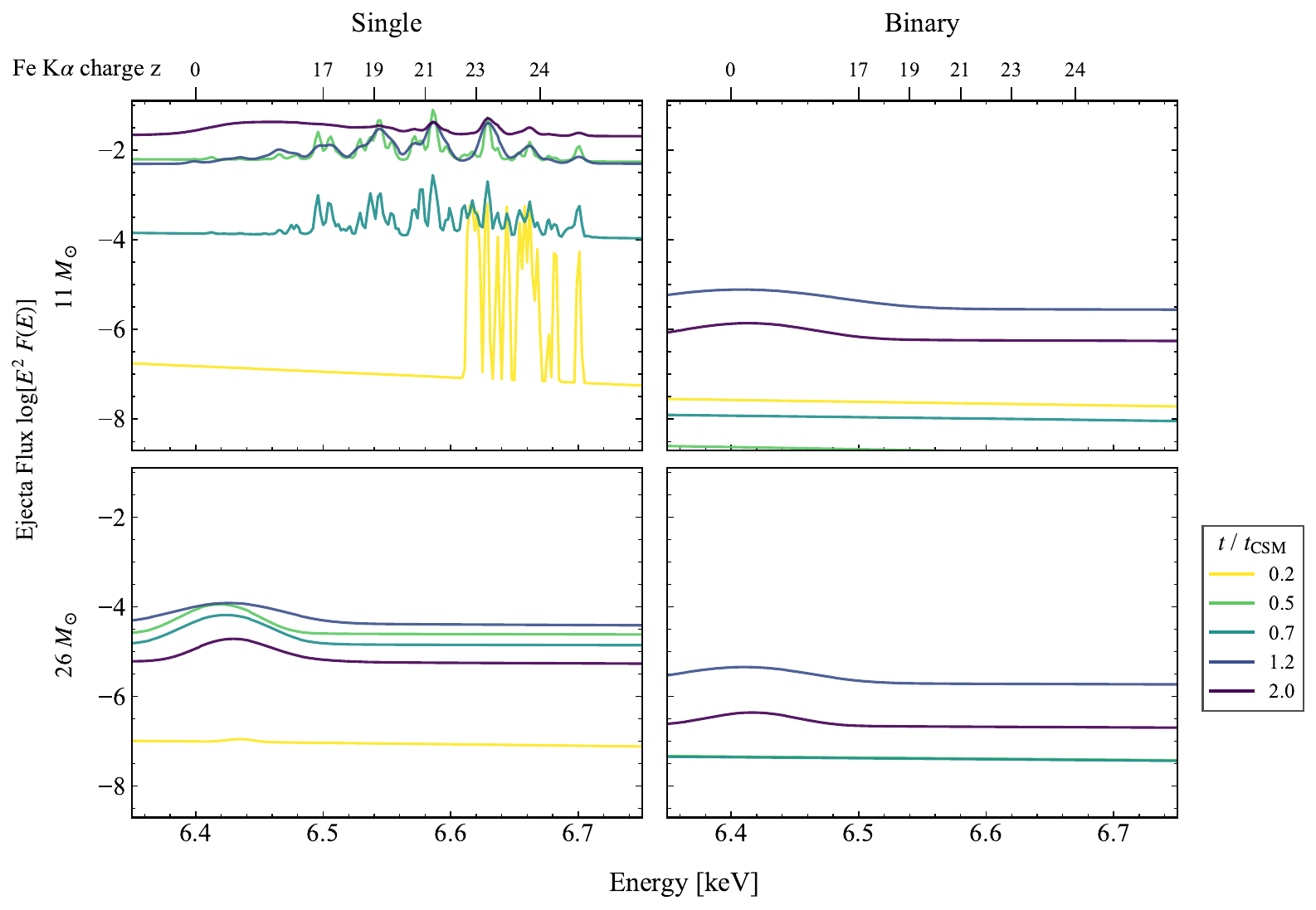}
\caption{Detailed view of the Fe line emission of our 11, 26 $M_{\odot}$ models.
The top and bottom panels show single and binary models; the left and right panels show the ejecta and CSM components, and the colors represent the time evolution scaled by $t_{\rm CSM}$.
Ticks on the top of upper panels represent the corresponding Fe K$\alpha$ energy.
The relatively narrow lines in single 11 $M_{\odot}$ model is a result of low ion temperature, and the absence of Doppler broadening.
In any case, as the observation is intrinsically spatially integrated, even XRISM is probably not capable of resolving such narrow lines \citep{Sapienza2024ProbingXRISM-Resolve}.
The figure shows that the emission from SESNRs exhibits a relatively low ionization state compared to "normal" CCSNRs (e.g., the Type IIP-like Single 11 $M_{\odot}$ model).}
\label{fig:snr_emission_fe}
\end{center}
\end{figure}
Figure~\ref{fig:snr_emission_fe} compares the Fe line emission.
The SESNR models exhibit relatively low luminosity due to the presence of the low-density bubble.
The single 11 $M_{\odot}$ model, likely of Type IIP SN origin, initially shows a relatively high ionization state, which then transitions to a lower state after the FS enters the dense shell ($t_{\rm CSM}\sim1$).
In contrast, the SESNR models do not show prominent Fe line emission until the FS hits the dense shell; moreover, even when the lines become prominent, the ionization state remain low.
This is because the high ionization state observed in the single 11 $M_{\odot}$ model results from the combination of high velocity, high ejecta density, and dense CSM.
By the time the FS of the SESNR models reache the dense shell, even though the FS retains considerable velocity, the low ejecta density resulting from the expansion of the low-mass ejecta prevents ions from reaching a high ionization state.
Since the SESNR models exhibit dim Fe line emission while inside the bubble and maintains low ionization even after becoming bright, our results caution against the simplified view that CCSNRs are uniquely characterized by high ionization states.
Instead, correct classification requires considering the line luminosity alongside the ionization state, reinforcing the diagnostic method presented by \citet{Yamaguchi2014DiscriminatingEmission}.
Moreover, our findings of the tendency in Fe line emission may serve as a criterion to distinguish SESNRs from other CCSNRs, since the features are consistent with SNRs thought to be such stripped progenitors \citep[e.g.,RX J1713.7-3946:][]{Katsuda2015EvidenceJ1713.7-3946}.

\begin{figure}[htbp]
\begin{center}
\includegraphics[width=180mm]{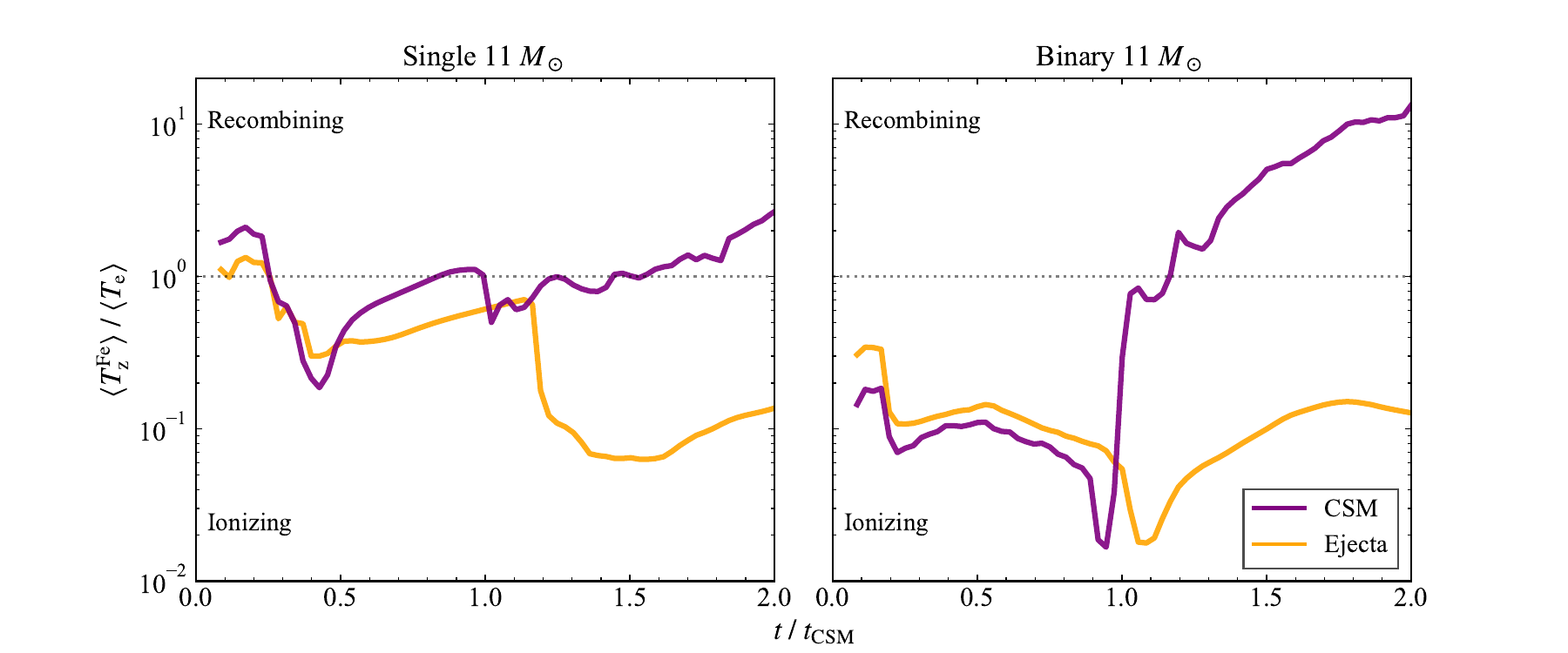}
\caption{Time evolution of the emission-measure-weighted average ionization temperature ratio, $\langle T_z \rangle / \langle T_e \rangle$, for Iron in the shocked ejecta (top panels) and shocked CSM (bottom panels) of our 11 $M_{\odot}$ models.
The x-axis is scaled by $t_{\rm CSM}$. A ratio $\ll 1$ indicates an ionizing plasma, $\gg 1$ indicates a recombining plasma, and $\sim 1$ indicates a plasma near collisional ionization equilibrium (CIE). 
The figure shows the correlation between the recombination features in the emission plots and the over-ionized phases.}
\label{fig:tzte_Fe}
\end{center}
\end{figure}

Figure~\ref{fig:tzte_Fe} shows the time evolution of the ionization temperature $T_z$ for Iron, divided by the electron temperature $T_e$.
The ionization temperature $T_z$ is the temperature that would produce the observed mean charge state of an element in collisional ionization equilibrium (CIE), as calculated using AtomDB \citep{Foster2023ATOMDBDatabase}.
The values are averaged for whole shocked region weighted by emission measure EM as follows:
\begin{align}
     \langle X \rangle &= \frac{\sum  EM_i X_i}{\sum EM_i}  \\
      EM_i &= n_{e,i} n_{{\rm ion},i} V_i
\end{align}
where $n_{e,i}$ and $n_{{\rm ion},i}$ represent the electron and total ion number densities in cell $i$, and $V_i$ is the volume of the cell.
The ratio $\langle T_z \rangle / \langle T_e \rangle$ indicates the ionization state of the plasma: a ratio of $\sim 1$ signifies a plasma in or near CIE;
a ratio $\ll 1$ indicates an under-ionized (ionizing) plasma; and a ratio $\gg 1$ indicates an over-ionized (recombining) plasma.
In the early stages, the binary 11 $M_{\odot}$ model shows an under-ionized plasma ($\langle T_z \rangle / \langle T_e \rangle \ll 1$) for both CSM and ejecta regions due to the low densities and high shock velocities which we discussed above.
On the other hand, the single 11 $M_{\odot}$ model shows an over-ionized plasma, which can exhibit recombining features due to its interaction with a denser RSG wind, a phenomenon also discussed in contexts like \citet{Moriya2012ProgenitorsRemnants} and confirmed in observations of some SNRs \citep[e.g.,][]{Yamaguchi2009DiscoverySuzaku}.
Conversely, after the forward shock hits the dense outer shell, both models show a significantly over-ionized CSM component ($\langle T_z \rangle / \langle T_e \rangle \gg 1$).
This indicates that the shocked CSM is cooling radiatively much faster than it can recombine.
These recombining features are visible as recombination edges in the X-ray spectra (Figure~\ref{fig:snr_emission}).

\section{DISCUSSIONS}\label{discussions}

\subsection{Comparison based on $M_{\rm ZAMS}$}
\subsubsection{$11 M_{\odot}$ Models}
The models with the lowest initial mass show the largest discrepancies in both dynamics and composition.
From the perspective of SN classification, our single $11 M_{\odot}$ model represents a classic Type IIP-like SNR, given its retained H/He envelope and the slow RSG wind prior to the explosion.
In contrast, our binary $11 M_{\odot}$ model represents a Type Ib-like SNR, as most of its H envelope has been stripped away by the time of the SN.
Since the binary $11 M_{\odot}$ model has a significantly lower ejecta mass ($1.82 M_{\odot}$), its initial shock velocity is nearly twice as high as that of its single counterpart.
Moreover, its pre-SN wind is faster ($\sim 500$ km s$^{-1}$), carving out a lower-density cavity.
Consequently, the FS velocity remains high ($\gtrsim 5000$ km s$^{-1}$) until it impacts the outer shell.
In this low-density environment, adiabatic cooling is highly efficient, keeping the average electron temperature in both the forward- and reverse-shocked plasma relatively low despite the high shock velocity.
In the single $11 M_{\odot}$ model, the denser CSM from the RSG wind leads to less efficient cooling and a higher ionization parameter, resulting in a higher average electron temperature.
The rapid decline and subsequent reheating observed in the single $11 M_{\odot}$ model signify the shock's interaction with the inner shell structure created during the progenitor's post-main-sequence evolution.
The low-density plasma of the binary $11 M_{\odot}$ model is also reflected in its emission features;
the continuum emission remains faint until the FS hits the outer shell.
As we noted, the binary $11 M_{\odot}$ model does not exhibit significant Fe line emission around 6.4-6.7 keV.
The low Fe abundance estimated from the observed SNR which is thought to be of binary origin due to its relatively low ejecta mass \citep{Katsuda2015EvidenceJ1713.7-3946}, is consistent with our result that binary 11 $M_{\odot}$ model shows less prominent Fe line emission.

\subsubsection{$26 M_{\odot}$ Models}
The medium initial mass models, which likely correspond to Type IIb (single) and Type Ib (binary) progenitors, also exhibit significant dynamical differences.
The binary $26 M_{\odot}$ model maintains a fast FS until hitting the outer shell, similar to the binary model of the lowest mass.
On the other hand, the single $26 M_{\odot}$ model possesses a more complex inner CSM with dense shells, a result of its "moderate" mass-loss history.
As the FS interacts with these inner structures, the RS can decelerate and decouple from the CD well before the FS reaches the main outer shell.
The evolution of the RS radius, a potentially observable quantity, could thus serve as a diagnostic for the progenitor's mass-loss history and, by extension, its single or binary nature.

\subsection{Comparison based on $M_{\rm ej}$}
From an observational perspective, the total ejecta mass ($M_{\rm ej}$) is a key parameter often inferred from SNR studies and is, in some respects, more important for comparing models than $M_{\rm ZAMS}$.
In our grid, the Single 11 $M_{\odot}$ and Binary 26 $M_{\odot}$ models, as well as the Single 26 $M_{\odot}$ and Binary 33 $M_{\odot}$ models, have similar $M_{\rm ej}$ values.
While these pairs of models with similar ejecta mass between single and binary origins have comparable initial shock velocities, their subsequent dynamical evolution and the chemical composition of their shocked ejecta are markedly different.
For instance, the lower-mass pair shows a difference in the prominence of shocked He core mass fraction (Figure~\ref{fig:snr_shockedejecta}), indicating the degree of exposure of the core due to the stripping.
This underscores the importance of using self-consistent progenitor models, as simply scaling a single-star model to match an inferred ejecta mass fails to capture the distinct evolutionary paths of binary systems.

\subsection{Caution to the interpretation of observation}\label{comparison}

\begin{figure}[htbp]
\begin{center}
\includegraphics[width=140mm]{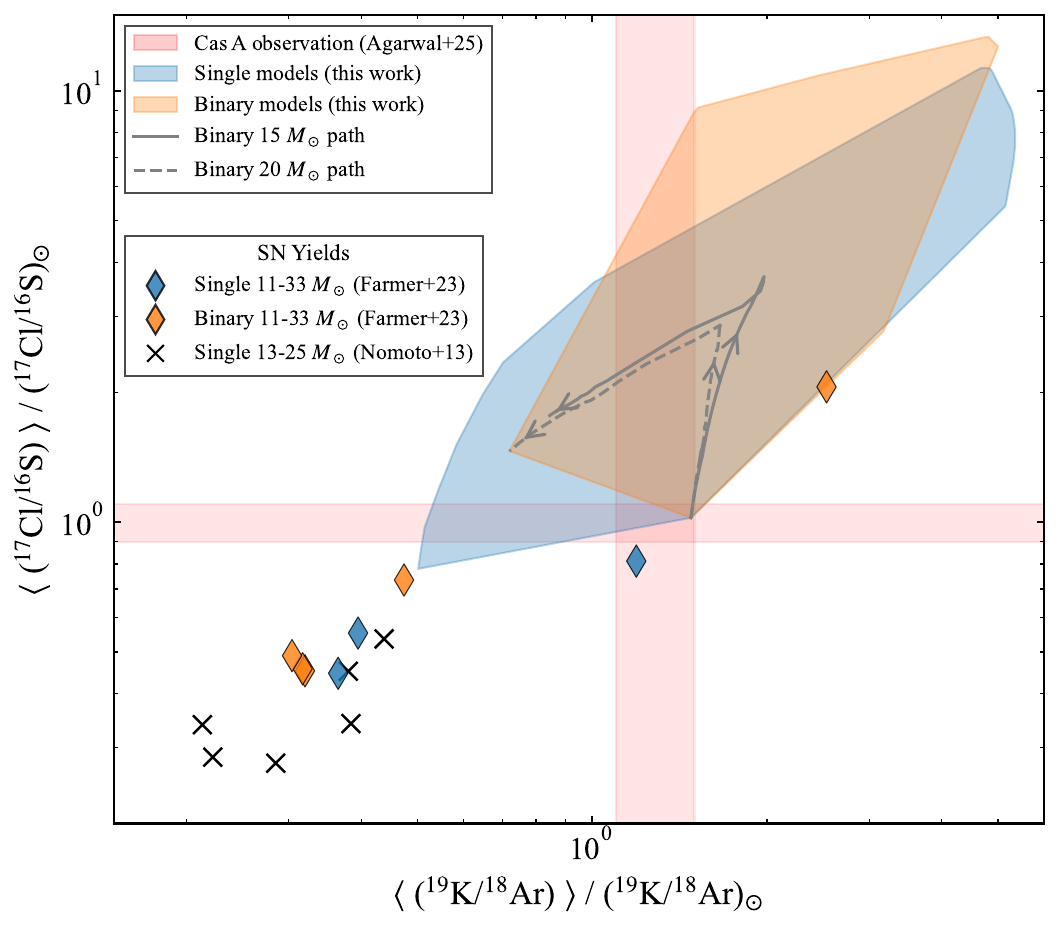}
\caption{Time evolution ($\sim 2.0$ $t_{\rm CSM}$) of Cl/S and K/Ar ratios of our models.
The blue/orange shaded areas represent the ranges of values covered by our single/binary models, while the red shaded area represents the observed ratios in the Cas A SE region \citep{Agarwal2025ChlorineRemnant}.
The values from our models are weighted by the emission measure of the numerator element, across shocked ejecta.
To compare with Cas A, we added time evolving track of binary 15 and 20 $M_{\odot}$ models with gray lines. 
Diamonds and crosses represent the SN yields: i.e., the direct comparison of SN ejecta vs. observation.
The figure indicates that comparing SNR observations directly to SN yield models should be done with caution; the SNR's dynamical evolution, in addition to the explosion mechanism, may be a key driver for the deviation from classical yields to the observed values (e.g., the enhanced odd-Z elements in Cas A).}
\label{fig:oddz}
\end{center}
\end{figure}

The lack of a convincing explanation for the origin of odd-Z elements remains a significant challenge in understanding Galactic chemical evolution.
Unprecedented high energy resolution spectroscopy with XRISM has recently detected such faint lines from odd-Z elements such as Cl and K from Cas A \citep{Agarwal2025ChlorineRemnant}.
In that study, by comparing the enhanced abundances of odd-Z elements in the SE region with the SN yields from several theoretical models, the observations suggest that Cas A has experienced pre-explosion activities, such as rotation, binarity, or shell merger, and classical SN models cannot reproduce the abundance pattern.
In this section, while we do not dispute this perspective, we highlight the necessity of exercising caution when directly comparing SNR emission to raw SN yields without accounting for SNR evolution.

Figure~\ref{fig:oddz} shows the time evolution of ejecta Cl/S and K/Ar abundance ratios of our models, compared with the observed Cas A abundance pattern \citep{Agarwal2025ChlorineRemnant}, and "pure" SN yields from theoretical nucleosynthesis models \citep{Nomoto2013NucleosynthesisGalaxies, Farmer2023NucleosynthesisStars}.
The odd-Z-ness $(^{17}\rm {Cl})/(^{16}\rm {S})$ and $(^{19}\rm {K})/(^{18}\rm {Ar})$, are taken from hydrodynamical output values in our simulation, weighted by $n_{\rm e} n_{\rm Cl,K} V$, to mimic the selection bias in observations where spatially resolved regions, such as the SE region of Cas A, are chosen for their brightness in these rare elements.

To specifically investigate the progenitor candidates for Cas A, we additionally calculated the binary 15 and 20 $M_{\odot}$ models.

All SNR models, including single star models, show the similar trends of rising odd-Z-ness at early phases of SNR evolution.
As the remnant evolves and the reverse shock penetrates deeper into the ejecta, these ratios eventually decrease, converging towards the values predicted by the pure SN yields.

These general trends suggest the importance of considering the SNR evolution, not only the pure SN yields. Using such yields to compare with SNR observations may be an oversimplification for comparing explosion models, as the line intensities of these rare elements are intrinsically sensitive to the dynamical state of the SNR.

Note that we do not focus on reproducing the exact observed Cas A value in our SNR models, since our simulation is 1-D and many studies suggest that to cover the dynamical and spectral features of Cas A, multi-dimensional calculations are needed.
However, again, our results highlight that numerical SNR calculations are key to bridge the gap between nucleosynthesis yields and observed emission patterns.

\section{CONCLUSIONS}\label{conclusions}
In this work, we have performed systematic, end-to-end hydrodynamical simulations of CCSNRs, consistently linking the pre-supernova evolution of single and binary progenitors to the subsequent remnant's dynamical and emission evolution.
Our primary findings are as follows:

\begin{enumerate}
    \item
    Progenitor binarity fundamentally alters SNR evolution.
    Mass loss via RLOF leads to lower ejecta masses and faster pre-supernova winds.
    This results in SNRs with higher initial shock velocities that expand into simpler, lower-density CSM cavities compared to their single-star counterparts.
    These differences are most pronounced for lower-mass progenitors (e.g., $11 M_{\odot}$).
    \item
    We have introduced a new characteristic timescale, $t_{\rm CSM}$, based on the total mass lost by the progenitor.
    This timescale successfully aligns the key evolutionary phases—expansion through the inner wind, the hot bubble, and interaction with the outer shell—across our diverse set of models.
    Though the traditional Sedov timescale, $t_{\rm sedov}$, is still a useful tool for estimating the time of RS reversal, our new $t_{\rm CSM}$ seems to be an additional key parameter for SNRs interacting with complex CSM environments.
    \item
    The distinct evolutionary paths of single and binary progenitors leave unique imprints on the X-ray properties of their SNRs.
    We find that the presence of prominent recombining features, indicating a rapidly cooling, over-ionized plasma, is a strong indicator of a shock interacting with a dense shell after traversing a low-density cavity—a scenario more common in our binary models.
    The temporal evolution of shocked ejecta composition and the relative brightness of ejecta versus CSM emission also provide clues to the progenitor's nature.
    \item
    Linking SNR emission to SN explosion mechanisms should be carried out carefully.
    We demonstrated that the observed odd-z-ness (Cl/S, K/Ar) is highly sensitive to the SNR evolution, and comparing pure SN yields directly may overlook the implication.
    Therefore, we underscore the critical importance of detailed SNR evolution modeling in bridging the gap between observations and theoretical nucleosynthesis models.
\end{enumerate}

Our study highlights the necessity of using self-consistent progenitor models to interpret the observed diversity of CCSNRs.
Simply scaling single-star models cannot capture the complex interplay between stellar evolution, mass loss, and SNR dynamics.
While our one-dimensional models have limitations (e.g., assuming spherical symmetry for RLOF and ignoring magnetic fields and particle acceleration), they provide a crucial theoretical foundation.
Future work should extend these models to two and three dimensions, incorporate a wider range of binary parameters, and include non-thermal processes to enable direct, multi-wavelength comparisons with observed SNRs.

\begin{acknowledgments}
   We thank Jacco Vink and Dan Milisavljevic for their helpful discussions and advice. We also thank Eva Laplace for kindly allowing us to use the stellar evolution models.
    \end{acknowledgments}

\newpage
\bibliography{references}{}
\bibliographystyle{aasjournalv7}

\appendix

\begin{deluxetable*}{c r r r r r r r r r}
\tablecaption{Evolution of Single Star Models \label{tab:single_evolution}}
\tablewidth{0pt}
\tablehead{
    \colhead{$M_{\rm ZAMS}$} &
    \colhead{$t/t_{\rm CSM}$} &
    \colhead{$t/t_{\rm sedov}$} &
    \colhead{$t$} &
    \colhead{$M^{\rm sh}_{\rm ej}$} &
    \colhead{$M^{\rm sh}_{\rm CSM}$} &
    \colhead{$\langle T_{\rm e} \rangle_{\rm ej}$} &
    \colhead{$\langle T_{\rm e} \rangle_{\rm CSM}$} &
    \colhead{$\langle n_{\rm e} t \rangle_{\rm ej}$} &
    \colhead{$\langle n_{\rm e} t \rangle_{\rm CSM}$} \\
    \colhead{($M_\odot$)} & \colhead{ } & \colhead{ } & \colhead{(yr)} & \colhead{($M_\odot$)} & \colhead{($M_\odot$)} & \colhead{(K)} & \colhead{(K)} & \colhead{(cm$^{-3}$ s)} & \colhead{(cm$^{-3}$ s)}
}
\startdata
11 & 0.087 & 0.078 & 153 & 0.11 & 0.04 & $2.0 \times 10^{6}$ & $5.7 \times 10^{6}$ & $3.4 \times 10^{11}$ & $1.2 \times 10^{11}$ \\
 & 0.200 & 0.180 & 353 & 0.17 & 0.07 & $1.2 \times 10^{6}$ & $4.4 \times 10^{6}$ & $3.1 \times 10^{11}$ & $1.1 \times 10^{11}$ \\
 & 0.313 & 0.282 & 553 & 0.25 & 0.15 & $3.2 \times 10^{6}$ & $8.3 \times 10^{6}$ & $2.6 \times 10^{11}$ & $7.1 \times 10^{10}$ \\
 & 0.398 & 0.359 & 704 & 0.37 & 0.71 & $2.1 \times 10^{7}$ & $7.0 \times 10^{6}$ & $2.0 \times 10^{11}$ & $6.8 \times 10^{9}$ \\
 & 0.511 & 0.460 & 904 & 1.44 & 1.29 & $1.5 \times 10^{7}$ & $8.6 \times 10^{6}$ & $9.6 \times 10^{10}$ & $1.6 \times 10^{10}$ \\
 & 0.596 & 0.537 & 1054 & 2.71 & 1.30 & $8.7 \times 10^{6}$ & $6.5 \times 10^{6}$ & $4.2 \times 10^{10}$ & $1.9 \times 10^{10}$ \\
 & 0.709 & 0.639 & 1254 & 3.79 & 1.31 & $6.5 \times 10^{6}$ & $5.2 \times 10^{6}$ & $3.1 \times 10^{10}$ & $2.1 \times 10^{10}$ \\
 & 0.794 & 0.715 & 1404 & 4.27 & 1.32 & $5.5 \times 10^{6}$ & $4.3 \times 10^{6}$ & $2.8 \times 10^{10}$ & $2.1 \times 10^{10}$ \\
 & 0.908 & 0.817 & 1604 & 4.71 & 1.34 & $4.5 \times 10^{6}$ & $3.4 \times 10^{6}$ & $2.5 \times 10^{10}$ & $2.1 \times 10^{10}$ \\
 & 0.993 & 0.894 & 1754 & 4.98 & 1.46 & $4.0 \times 10^{6}$ & $3.0 \times 10^{6}$ & $2.5 \times 10^{10}$ & $1.9 \times 10^{10}$ \\
 & 1.106 & 0.996 & 1954 & 5.24 & 6.31 & $3.4 \times 10^{6}$ & $1.9 \times 10^{6}$ & $2.4 \times 10^{10}$ & $3.4 \times 10^{10}$ \\
 & 1.502 & 1.353 & 2655 & 5.75 & 179.77 & $6.3 \times 10^{7}$ & $1.3 \times 10^{5}$ & $3.3 \times 10^{10}$ & $1.3 \times 10^{12}$ \\
 & 2.012 & 1.812 & 3556 & 7.12 & 309.28 & $5.2 \times 10^{7}$ & $7.5 \times 10^{3}$ & $3.6 \times 10^{10}$ & $1.6 \times 10^{13}$ \\
\tableline
26 & 0.098 & 0.226 & 804 & 0.05 & 0.03 & $1.6 \times 10^{6}$ & $1.6 \times 10^{6}$ & $4.6 \times 10^{9}$ & $2.1 \times 10^{9}$ \\
 & 0.201 & 0.466 & 1654 & 0.09 & 0.08 & $3.5 \times 10^{6}$ & $5.8 \times 10^{6}$ & $4.6 \times 10^{9}$ & $1.5 \times 10^{9}$ \\
 & 0.299 & 0.692 & 2455 & 0.23 & 0.24 & $5.5 \times 10^{6}$ & $2.8 \times 10^{6}$ & $2.4 \times 10^{9}$ & $8.1 \times 10^{8}$ \\
 & 0.403 & 0.931 & 3306 & 0.56 & 0.54 & $5.6 \times 10^{6}$ & $1.7 \times 10^{6}$ & $2.3 \times 10^{9}$ & $4.3 \times 10^{8}$ \\
 & 0.500 & 1.157 & 4106 & 1.02 & 12.01 & $1.8 \times 10^{7}$ & $3.8 \times 10^{5}$ & $2.0 \times 10^{9}$ & $6.2 \times 10^{10}$ \\
 & 0.598 & 1.383 & 4907 & 3.50 & 12.11 & $9.4 \times 10^{6}$ & $6.5 \times 10^{5}$ & $9.7 \times 10^{8}$ & $1.9 \times 10^{11}$ \\
 & 0.701 & 1.622 & 5757 & 5.79 & 12.36 & $6.2 \times 10^{6}$ & $7.8 \times 10^{5}$ & $9.6 \times 10^{8}$ & $2.8 \times 10^{11}$ \\
 & 0.799 & 1.848 & 6558 & 7.17 & 12.65 & $5.4 \times 10^{6}$ & $7.2 \times 10^{5}$ & $1.1 \times 10^{9}$ & $3.1 \times 10^{11}$ \\
 & 0.902 & 2.088 & 7409 & 8.04 & 13.00 & $4.7 \times 10^{6}$ & $5.9 \times 10^{5}$ & $1.3 \times 10^{9}$ & $3.4 \times 10^{11}$ \\
 & 1.000 & 2.313 & 8209 & 8.59 & 13.44 & $4.1 \times 10^{6}$ & $4.6 \times 10^{5}$ & $1.4 \times 10^{9}$ & $3.7 \times 10^{11}$ \\
 & 1.097 & 2.539 & 9010 & 8.96 & 50.17 & $3.7 \times 10^{6}$ & $2.1 \times 10^{6}$ & $1.4 \times 10^{9}$ & $4.7 \times 10^{10}$ \\
 & 1.499 & 3.470 & 12312 & 9.62 & 1361.47 & $1.2 \times 10^{7}$ & $5.3 \times 10^{3}$ & $2.1 \times 10^{9}$ & $6.0 \times 10^{12}$ \\
 & 1.999 & 4.626 & 16416 & 9.86 & 1809.49 & $1.0 \times 10^{7}$ & $4.8 \times 10^{3}$ & $2.5 \times 10^{9}$ & $8.8 \times 10^{12}$ \\
\tableline
33 & 0.098 & 0.138 & 1054 & 0.04 & 0.03 & $1.3 \times 10^{6}$ & $1.2 \times 10^{6}$ & $3.1 \times 10^{9}$ & $1.6 \times 10^{9}$ \\
 & 0.201 & 0.282 & 2155 & 0.08 & 0.06 & $1.0 \times 10^{6}$ & $1.0 \times 10^{6}$ & $2.8 \times 10^{9}$ & $1.6 \times 10^{9}$ \\
 & 0.298 & 0.419 & 3205 & 0.10 & 0.09 & $9.5 \times 10^{5}$ & $7.8 \times 10^{5}$ & $2.8 \times 10^{9}$ & $1.5 \times 10^{9}$ \\
 & 0.401 & 0.563 & 4306 & 0.18 & 0.29 & $5.9 \times 10^{6}$ & $3.7 \times 10^{6}$ & $1.9 \times 10^{9}$ & $8.5 \times 10^{8}$ \\
 & 0.499 & 0.700 & 5357 & 0.45 & 0.62 & $4.4 \times 10^{6}$ & $2.8 \times 10^{6}$ & $1.3 \times 10^{9}$ & $6.4 \times 10^{8}$ \\
 & 0.601 & 0.844 & 6458 & 0.81 & 1.40 & $3.3 \times 10^{6}$ & $6.2 \times 10^{5}$ & $8.8 \times 10^{8}$ & $1.0 \times 10^{8}$ \\
 & 0.699 & 0.981 & 7509 & 1.32 & 10.04 & $4.0 \times 10^{6}$ & $1.0 \times 10^{6}$ & $6.7 \times 10^{8}$ & $1.1 \times 10^{9}$ \\
 & 0.801 & 1.125 & 8610 & 3.41 & 15.70 & $3.5 \times 10^{6}$ & $2.9 \times 10^{6}$ & $2.6 \times 10^{8}$ & $9.0 \times 10^{9}$ \\
 & 0.899 & 1.263 & 9660 & 6.00 & 16.50 & $2.7 \times 10^{6}$ & $2.5 \times 10^{6}$ & $2.7 \times 10^{8}$ & $1.9 \times 10^{10}$ \\
 & 1.002 & 1.407 & 10761 & 8.18 & 18.86 & $2.8 \times 10^{6}$ & $1.9 \times 10^{6}$ & $3.9 \times 10^{8}$ & $2.9 \times 10^{10}$ \\
 & 1.095 & 1.537 & 11762 & 9.58 & 22.84 & $2.7 \times 10^{6}$ & $3.2 \times 10^{6}$ & $4.9 \times 10^{8}$ & $3.3 \times 10^{10}$ \\
 & 1.491 & 2.093 & 16015 & 11.79 & 903.39 & $5.7 \times 10^{6}$ & $5.7 \times 10^{3}$ & $10.0 \times 10^{8}$ & $1.6 \times 10^{12}$ \\
 & 1.999 & 2.806 & 21469 & 12.33 & 1818.32 & $5.0 \times 10^{6}$ & $2.2 \times 10^{3}$ & $1.4 \times 10^{9}$ & $4.2 \times 10^{12}$ \\
\enddata
\tablecomments{$t_{\rm CSM}$ is the age when $M^{\rm sh}_{\rm CSM} = M^{\rm loss}_{\rm CSM}$. $t_{\rm sedov}$ is the age when $M^{\rm sh}_{\rm CSM} = M^{\rm init}_{\rm ej}$. The values of $(t_{\rm CSM}, t_{\rm sedov})$ (yr) for each model are: $M_{\rm ZAMS}=11$: $(1767, 1962)$, $M_{\rm ZAMS}=26$: $(8212, 3549)$, $M_{\rm ZAMS}=33$: $(10742, 7650)$.}
\end{deluxetable*}

\begin{deluxetable*}{c r r r r r r r r r}
\tablecaption{Evolution of Binary Star Models \label{tab:binary_evolution}}
\tablewidth{0pt}
\tablehead{
    \colhead{$M_{\rm ZAMS}$} &
    \colhead{$t/t_{\rm CSM}$} &
    \colhead{$t/t_{\rm sedov}$} &
    \colhead{$t$} &
    \colhead{$M^{\rm sh}_{\rm ej}$} &
    \colhead{$M^{\rm sh}_{\rm CSM}$} &
    \colhead{$\langle T_{\rm e} \rangle_{\rm ej}$} &
    \colhead{$\langle T_{\rm e} \rangle_{\rm CSM}$} &
    \colhead{$\langle n_{\rm e} t \rangle_{\rm ej}$} &
    \colhead{$\langle n_{\rm e} t \rangle_{\rm CSM}$} \\
    \colhead{($M_\odot$)} & \colhead{ } & \colhead{ } & \colhead{(yr)} & \colhead{($M_\odot$)} & \colhead{($M_\odot$)} & \colhead{(K)} & \colhead{(K)} & \colhead{(cm$^{-3}$ s)} & \colhead{(cm$^{-3}$ s)}
}
\startdata
11 & 0.113 & 0.115 & 203 & 0.02 & 0.01 & $1.9 \times 10^{6}$ & $1.4 \times 10^{6}$ & $1.2 \times 10^{9}$ & $3.0 \times 10^{8}$ \\
 & 0.196 & 0.200 & 353 & 0.02 & 0.01 & $6.5 \times 10^{6}$ & $3.1 \times 10^{6}$ & $1.3 \times 10^{9}$ & $2.8 \times 10^{8}$ \\
 & 0.307 & 0.313 & 553 & 0.05 & 0.03 & $3.9 \times 10^{6}$ & $2.4 \times 10^{6}$ & $7.8 \times 10^{8}$ & $2.3 \times 10^{8}$ \\
 & 0.390 & 0.398 & 704 & 0.09 & 0.05 & $3.3 \times 10^{6}$ & $2.0 \times 10^{6}$ & $7.1 \times 10^{8}$ & $2.3 \times 10^{8}$ \\
 & 0.501 & 0.511 & 904 & 0.13 & 0.07 & $2.5 \times 10^{6}$ & $1.6 \times 10^{6}$ & $6.2 \times 10^{8}$ & $1.9 \times 10^{8}$ \\
 & 0.612 & 0.624 & 1104 & 0.18 & 0.10 & $2.7 \times 10^{6}$ & $1.6 \times 10^{6}$ & $6.0 \times 10^{8}$ & $1.7 \times 10^{8}$ \\
 & 0.696 & 0.709 & 1254 & 0.21 & 0.12 & $2.9 \times 10^{6}$ & $1.6 \times 10^{6}$ & $5.4 \times 10^{8}$ & $1.3 \times 10^{8}$ \\
 & 0.807 & 0.822 & 1454 & 0.24 & 0.17 & $2.6 \times 10^{6}$ & $1.2 \times 10^{6}$ & $4.0 \times 10^{8}$ & $7.1 \times 10^{7}$ \\
 & 0.890 & 0.907 & 1604 & 0.28 & 0.23 & $2.6 \times 10^{6}$ & $7.7 \times 10^{5}$ & $3.5 \times 10^{8}$ & $2.8 \times 10^{7}$ \\
 & 1.001 & 1.020 & 1804 & 0.37 & 4.41 & $2.7 \times 10^{6}$ & $1.1 \times 10^{6}$ & $2.6 \times 10^{8}$ & $1.0 \times 10^{9}$ \\
 & 1.112 & 1.133 & 2005 & 0.58 & 47.69 & $5.7 \times 10^{6}$ & $5.8 \times 10^{5}$ & $1.4 \times 10^{8}$ & $8.7 \times 10^{10}$ \\
 & 1.501 & 1.529 & 2705 & 1.58 & 284.24 & $4.3 \times 10^{6}$ & $4.2 \times 10^{3}$ & $4.8 \times 10^{8}$ & $1.9 \times 10^{12}$ \\
 & 2.001 & 2.038 & 3606 & 1.80 & 607.94 & $5.0 \times 10^{6}$ & $1.4 \times 10^{3}$ & $8.7 \times 10^{8}$ & $5.9 \times 10^{12}$ \\
\tableline
15 & 0.096 & 0.097 & 303 & 0.02 & 0.01 & $1.5 \times 10^{6}$ & $1.4 \times 10^{6}$ & $7.2 \times 10^{8}$ & $3.3 \times 10^{8}$ \\
 & 0.207 & 0.210 & 653 & 0.03 & 0.02 & $1.0 \times 10^{6}$ & $8.5 \times 10^{5}$ & $7.6 \times 10^{8}$ & $3.1 \times 10^{8}$ \\
 & 0.302 & 0.306 & 954 & 0.06 & 0.04 & $3.3 \times 10^{6}$ & $2.3 \times 10^{6}$ & $5.2 \times 10^{8}$ & $2.2 \times 10^{8}$ \\
 & 0.397 & 0.402 & 1254 & 0.10 & 0.08 & $2.9 \times 10^{6}$ & $1.8 \times 10^{6}$ & $4.5 \times 10^{8}$ & $1.6 \times 10^{8}$ \\
 & 0.492 & 0.499 & 1554 & 0.17 & 0.13 & $2.6 \times 10^{6}$ & $1.4 \times 10^{6}$ & $3.5 \times 10^{8}$ & $1.1 \times 10^{8}$ \\
 & 0.603 & 0.611 & 1904 & 0.28 & 0.21 & $2.2 \times 10^{6}$ & $1.1 \times 10^{6}$ & $2.6 \times 10^{8}$ & $6.9 \times 10^{7}$ \\
 & 0.699 & 0.707 & 2205 & 0.39 & 0.30 & $1.9 \times 10^{6}$ & $9.0 \times 10^{5}$ & $2.0 \times 10^{8}$ & $4.7 \times 10^{7}$ \\
 & 0.794 & 0.804 & 2505 & 0.52 & 0.41 & $1.7 \times 10^{6}$ & $7.6 \times 10^{5}$ & $1.6 \times 10^{8}$ & $3.3 \times 10^{7}$ \\
 & 0.905 & 0.916 & 2855 & 0.71 & 0.57 & $1.3 \times 10^{6}$ & $5.7 \times 10^{5}$ & $1.1 \times 10^{8}$ & $1.7 \times 10^{7}$ \\
 & 1.000 & 1.013 & 3155 & 0.91 & 5.76 & $1.3 \times 10^{6}$ & $1.1 \times 10^{6}$ & $9.5 \times 10^{7}$ & $9.1 \times 10^{8}$ \\
 & 1.095 & 1.109 & 3456 & 1.07 & 75.49 & $3.0 \times 10^{6}$ & $2.5 \times 10^{5}$ & $7.8 \times 10^{7}$ & $4.6 \times 10^{10}$ \\
 & 1.507 & 1.526 & 4757 & 2.74 & 381.75 & $2.6 \times 10^{6}$ & $3.3 \times 10^{3}$ & $2.5 \times 10^{8}$ & $1.8 \times 10^{12}$ \\
 & 1.999 & 2.024 & 6308 & 3.14 & 743.41 & $3.5 \times 10^{6}$ & $1.1 \times 10^{3}$ & $4.9 \times 10^{8}$ & $4.5 \times 10^{12}$ \\
\tableline
20 & 0.106 & 0.106 & 403 & 0.02 & 0.01 & $1.4 \times 10^{6}$ & $1.3 \times 10^{6}$ & $1.4 \times 10^{9}$ & $7.5 \times 10^{8}$ \\
 & 0.197 & 0.198 & 754 & 0.03 & 0.02 & $1.2 \times 10^{6}$ & $1.1 \times 10^{6}$ & $1.3 \times 10^{9}$ & $7.0 \times 10^{8}$ \\
 & 0.302 & 0.303 & 1154 & 0.06 & 0.06 & $5.3 \times 10^{6}$ & $3.2 \times 10^{6}$ & $8.6 \times 10^{8}$ & $3.9 \times 10^{8}$ \\
 & 0.394 & 0.395 & 1504 & 0.13 & 0.12 & $4.0 \times 10^{6}$ & $2.1 \times 10^{6}$ & $6.7 \times 10^{8}$ & $2.5 \times 10^{8}$ \\
 & 0.499 & 0.500 & 1904 & 0.24 & 0.22 & $3.1 \times 10^{6}$ & $1.4 \times 10^{6}$ & $4.8 \times 10^{8}$ & $1.5 \times 10^{8}$ \\
 & 0.603 & 0.606 & 2305 & 0.40 & 0.34 & $2.4 \times 10^{6}$ & $1.0 \times 10^{6}$ & $3.5 \times 10^{8}$ & $1.0 \times 10^{8}$ \\
 & 0.695 & 0.698 & 2655 & 0.57 & 0.48 & $2.0 \times 10^{6}$ & $8.4 \times 10^{5}$ & $2.8 \times 10^{8}$ & $7.4 \times 10^{7}$ \\
 & 0.800 & 0.803 & 3055 & 0.82 & 0.67 & $1.5 \times 10^{6}$ & $6.1 \times 10^{5}$ & $2.0 \times 10^{8}$ & $4.3 \times 10^{7}$ \\
 & 0.905 & 0.908 & 3456 & 1.13 & 1.44 & $1.3 \times 10^{6}$ & $3.0 \times 10^{5}$ & $1.4 \times 10^{8}$ & $3.6 \times 10^{7}$ \\
 & 0.996 & 1.000 & 3806 & 1.47 & 5.42 & $9.2 \times 10^{5}$ & $1.1 \times 10^{6}$ & $9.7 \times 10^{7}$ & $5.7 \times 10^{8}$ \\
 & 1.101 & 1.105 & 4206 & 1.86 & 98.28 & $2.0 \times 10^{6}$ & $2.1 \times 10^{5}$ & $1.1 \times 10^{8}$ & $7.2 \times 10^{10}$ \\
 & 1.494 & 1.500 & 5707 & 4.62 & 469.70 & $4.8 \times 10^{6}$ & $2.7 \times 10^{3}$ & $1.1 \times 10^{9}$ & $2.4 \times 10^{12}$ \\
 & 2.005 & 2.012 & 7659 & 5.42 & 762.89 & $5.8 \times 10^{6}$ & $1.4 \times 10^{3}$ & $2.0 \times 10^{9}$ & $7.8 \times 10^{12}$ \\
\enddata
\tablecomments{$t_{\rm CSM}$ is the age when $M^{\rm sh}_{\rm CSM} = M^{\rm loss}_{\rm CSM}$. $t_{\rm sedov}$ is the age when $M^{\rm sh}_{\rm CSM} = M^{\rm init}_{\rm ej}$. The values of $(t_{\rm CSM}, t_{\rm sedov})$ (yr) for each model are: $M_{\rm ZAMS}=11$: $(1802, 1769)$, $M_{\rm ZAMS}=15$: $(3156, 3116)$, $M_{\rm ZAMS}=20$: $(3820, 3806)$, $M_{\rm ZAMS}=26$: $(5864, 5861)$, $M_{\rm ZAMS}=33$: $(8765, 8762)$.}
\end{deluxetable*}

\begin{deluxetable*}{c r r r r r r r r r}
\tablecaption{Evolution of Binary Star Models (Cont.)} 
\tablewidth{0pt}
\tablehead{
    \colhead{$M_{\rm ZAMS}$} &
    \colhead{$t/t_{\rm CSM}$} &
    \colhead{$t/t_{\rm sedov}$} &
    \colhead{$t$} &
    \colhead{$M^{\rm sh}_{\rm ej}$} &
    \colhead{$M^{\rm sh}_{\rm CSM}$} &
    \colhead{$\langle T_{\rm e} \rangle_{\rm ej}$} &
    \colhead{$\langle T_{\rm e} \rangle_{\rm CSM}$} &
    \colhead{$\langle n_{\rm e} t \rangle_{\rm ej}$} &
    \colhead{$\langle n_{\rm e} t \rangle_{\rm CSM}$} \\
    \colhead{($M_\odot$)} & \colhead{ } & \colhead{ } & \colhead{(yr)} & \colhead{($M_\odot$)} & \colhead{($M_\odot$)} & \colhead{(K)} & \colhead{(K)} & \colhead{(cm$^{-3}$ s)} & \colhead{(cm$^{-3}$ s)}
}
\startdata
26 & 0.103 & 0.103 & 603 & 0.02 & 0.01 & $1.4 \times 10^{6}$ & $1.1 \times 10^{6}$ & $1.4 \times 10^{9}$ & $7.4 \times 10^{8}$ \\
 & 0.197 & 0.197 & 1154 & 0.04 & 0.04 & $1.2 \times 10^{6}$ & $6.6 \times 10^{5}$ & $1.3 \times 10^{9}$ & $5.6 \times 10^{8}$ \\
 & 0.299 & 0.299 & 1754 & 0.12 & 0.13 & $4.9 \times 10^{6}$ & $1.8 \times 10^{6}$ & $7.3 \times 10^{8}$ & $2.2 \times 10^{8}$ \\
 & 0.402 & 0.402 & 2355 & 0.28 & 0.29 & $3.6 \times 10^{6}$ & $1.2 \times 10^{6}$ & $4.6 \times 10^{8}$ & $1.4 \times 10^{8}$ \\
 & 0.504 & 0.504 & 2955 & 0.54 & 0.48 & $2.5 \times 10^{6}$ & $8.4 \times 10^{5}$ & $3.1 \times 10^{8}$ & $9.0 \times 10^{7}$ \\
 & 0.598 & 0.598 & 3506 & 0.83 & 0.71 & $1.9 \times 10^{6}$ & $6.6 \times 10^{5}$ & $2.2 \times 10^{8}$ & $7.0 \times 10^{7}$ \\
 & 0.700 & 0.701 & 4106 & 1.24 & 1.02 & $1.4 \times 10^{6}$ & $5.6 \times 10^{5}$ & $1.4 \times 10^{8}$ & $5.5 \times 10^{7}$ \\
 & 0.803 & 0.803 & 4707 & 1.89 & 1.40 & $1.1 \times 10^{6}$ & $4.8 \times 10^{5}$ & $9.8 \times 10^{7}$ & $4.1 \times 10^{7}$ \\
 & 0.896 & 0.897 & 5257 & 2.31 & 1.76 & $1.3 \times 10^{6}$ & $4.7 \times 10^{5}$ & $1.2 \times 10^{8}$ & $3.5 \times 10^{7}$ \\
 & 0.999 & 0.999 & 5857 & 2.72 & 2.72 & $1.3 \times 10^{6}$ & $1.5 \times 10^{5}$ & $1.6 \times 10^{8}$ & $3.8 \times 10^{6}$ \\
 & 1.101 & 1.102 & 6458 & 3.23 & 80.59 & $1.4 \times 10^{6}$ & $1.5 \times 10^{5}$ & $2.2 \times 10^{8}$ & $6.9 \times 10^{10}$ \\
 & 1.511 & 1.512 & 8860 & 5.97 & 526.65 & $4.9 \times 10^{6}$ & $2.1 \times 10^{3}$ & $9.0 \times 10^{8}$ & $2.0 \times 10^{12}$ \\
 & 1.997 & 1.998 & 11712 & 7.26 & 1222.65 & $5.0 \times 10^{6}$ & $5.5 \times 10^{2}$ & $1.4 \times 10^{9}$ & $5.5 \times 10^{12}$ \\
\tableline
33 & 0.097 & 0.097 & 854 & 0.03 & 0.02 & $1.3 \times 10^{6}$ & $1.0 \times 10^{6}$ & $1.1 \times 10^{9}$ & $5.6 \times 10^{8}$ \\
 & 0.200 & 0.200 & 1754 & 0.05 & 0.04 & $1.1 \times 10^{6}$ & $8.6 \times 10^{5}$ & $1.0 \times 10^{9}$ & $5.3 \times 10^{8}$ \\
 & 0.303 & 0.303 & 2655 & 0.15 & 0.22 & $4.9 \times 10^{6}$ & $1.5 \times 10^{6}$ & $5.1 \times 10^{8}$ & $1.1 \times 10^{8}$ \\
 & 0.400 & 0.400 & 3506 & 0.42 & 0.47 & $3.0 \times 10^{6}$ & $7.4 \times 10^{5}$ & $3.0 \times 10^{8}$ & $5.0 \times 10^{7}$ \\
 & 0.503 & 0.503 & 4406 & 0.82 & 0.82 & $1.9 \times 10^{6}$ & $4.7 \times 10^{5}$ & $2.3 \times 10^{8}$ & $3.6 \times 10^{7}$ \\
 & 0.600 & 0.600 & 5257 & 1.31 & 1.25 & $1.4 \times 10^{6}$ & $4.0 \times 10^{5}$ & $1.9 \times 10^{8}$ & $4.6 \times 10^{7}$ \\
 & 0.703 & 0.703 & 6158 & 2.14 & 1.83 & $9.4 \times 10^{5}$ & $3.9 \times 10^{5}$ & $6.6 \times 10^{7}$ & $2.8 \times 10^{7}$ \\
 & 0.800 & 0.800 & 7008 & 3.06 & 2.48 & $8.5 \times 10^{5}$ & $3.7 \times 10^{5}$ & $6.4 \times 10^{7}$ & $1.8 \times 10^{7}$ \\
 & 0.902 & 0.903 & 7909 & 3.80 & 3.30 & $1.1 \times 10^{6}$ & $3.0 \times 10^{5}$ & $9.5 \times 10^{7}$ & $1.1 \times 10^{7}$ \\
 & 0.999 & 1.000 & 8760 & 4.32 & 4.46 & $1.2 \times 10^{6}$ & $1.8 \times 10^{5}$ & $1.4 \times 10^{8}$ & $4.6 \times 10^{6}$ \\
 & 1.114 & 1.114 & 9760 & 4.97 & 152.96 & $1.4 \times 10^{6}$ & $6.2 \times 10^{4}$ & $2.0 \times 10^{8}$ & $1.2 \times 10^{11}$ \\
 & 1.513 & 1.514 & 13263 & 7.98 & 852.56 & $4.1 \times 10^{6}$ & $2.0 \times 10^{3}$ & $7.1 \times 10^{8}$ & $2.1 \times 10^{12}$ \\
 & 1.999 & 1.999 & 17516 & 9.70 & 1938.41 & $4.1 \times 10^{6}$ & $4.1 \times 10^{2}$ & $1.1 \times 10^{9}$ & $4.6 \times 10^{12}$ \\
\enddata
\end{deluxetable*}

We summarize the key SNR properties at our reference times in Tables~\ref{tab:single_evolution} and \ref{tab:binary_evolution}, including shocked mass, average electron temperatures, and ionization timescales for both ejecta and CSM.
To facilitate comparison with observational data, the plasma state values ($\langle T_{\rm e} \rangle$, $\langle n_{\rm e} t \rangle$) are weighted by the emission measure ($n_{\rm e} n_{\rm ion} V$, where $n_{\rm e}$ is the electron density, $n_{\rm ion}$ is the average ion density, and $V$ is the volume of each grid cell).
The table suggests that binary models possess relatively low-temperature plasma due to the low-density CSM. This low density maintains high FS velocities, causing adiabatic cooling to dominate over collisional heating.

\end{document}